\documentclass[10pt]{article}
\pdfoutput=1
\usepackage[utf8]{inputenc}
\usepackage{cite}
\usepackage{amsmath,amssymb,amsbsy,amstext,amsthm,simplewick,amsfonts}
\usepackage{graphicx}
\usepackage{subcaption}
\usepackage{lscape}
\usepackage{cancel}
\usepackage{wrapfig}
\usepackage{upgreek}
\usepackage{bm} 
\usepackage{framed}
\usepackage{bbm}
\usepackage{textcomp}
\usepackage{tikz}
\usepackage{pifont}
\usetikzlibrary{matrix,shapes,fit,tikzmark,calc}
\usepackage{adjustbox}
\usepackage{makecell}
\usepackage{tcolorbox}
\usepackage{physics}
\usepackage{empheq}
\usepackage[normalem]{ulem}
\usepackage{enumitem}
\usepackage{braket} 
\usepackage{array}
\usepackage{dsfont}
\usepackage{ulem}
\usepackage{multirow}
\usepackage{titlesec}
\titleformat{\section}{\normalfont\fontsize{12}{16}\bfseries}{\thesection}{1em}{}

\numberwithin{equation}{section}

\def\be{\begin{equation}}
\def\ee{\end{equation}}

\def\ba{\begin{eqnarray}}
\def\ea{\end{eqnarray}}

\def\bfk{\textbf{k}}

\newcommand{\Disc}[2]{\underset{#2}{\text{Disc\,}}#1}

\definecolor{blue3}{RGB}{31,119,180}
\definecolor{red3}{RGB}{214,39,40}
\definecolor{orange3}{RGB}{255,127,14}
\definecolor{green3}{RGB}{44,160,44}

\usepackage{colortbl}
\definecolor{lightgreen}{cmyk}{0.2, 0, 0.2, 0.2}
\definecolor{lightgray}{cmyk}{0.1,0.2,0,0.1}
\definecolor{lightgray2}{cmyk}{0.1,0.1,0,0.1}

\usepackage{hyperref}
\hypersetup{colorlinks=true,linkcolor=teal,citecolor=orange3,urlcolor=green3,pdfencoding=auto}

\makeatletter
\newlength{\apb@width}
\newcommand{\autoparbox}[2][c]{\settowidth{\apb@width}{#2}\parbox[#1]{\apb@width}{#2}}

\makeatother


\def\bfp{\textbf{p}}

\def\beq{\begin{equation}}
\def\eeq{\end{equation}}

\newcommand{\dt}{\tilde \delta_{D}^{(3)}}

\begin{document}
\begin{center}


{\fontsize
{20}{20} \bf From Amplitudes to Analytic Wavefunctions} \\

\end{center}

\vskip 18pt
\begin{center}
\noindent
{\fontsize{12}{18}\selectfont Mang Hei Gordon Lee\footnote{\tt  mhgl2@cam.ac.uk}}
\end{center}

\begin{center}
\vskip 8pt
\textit{Department of Applied Mathematics and Theoretical Physics, University of Cambridge, Wilberforce Road, Cambridge, CB3 0WA, UK} 
\end{center}
\begin{abstract}
The field-theoretic wavefunction has received renewed attention with the goal of better understanding observables at the boundary of de Sitter spacetime and studying the interior of Minkowski or general FLRW spacetime. Understanding the analytic structure of the wavefunction potentially allows us to establish bounds on physical observables. In this paper we develop an "amplitude representation" for the flat space wavefunction, which allow us to write the flat space wavefunction as an amplitude-like Feynman integral integrated over an energy-fixing kernel. With this representation it is possible to separate the wavefunction into an amplitude part and a subleading part which is less divergent as the total energy goes to zero. In turn the singularities of the wavefunction can be classified into two sets: amplitude-type singularities, which can be mapped to singularities found in amplitudes (including anomalous thresholds), and wavefunction-type singularities, which are unique to the wavefunction. As an example we study several tree level and one loop diagrams for scalars, and explore their singularities in detail.
\end{abstract}


\section{Introduction}
Analyticity has played a crucial role in understanding scattering amplitudes in quantum field theory. By considering general principles of the underlying physical theory, such as locality, causality and crossing, it can be shown that a scattering amplitude must have a very specific analytic structure. The details of the analyticity of amplitude can often tell us more about the theory, such as the mass and spin of states present in the theory \cite{Eden:1966dnq, Martin:1969ina, Benincasa:2013, Elvang:2015rqa, Cheung:2017pzi}.

Understanding the analytic structure of amplitudes also leads to useful constraints. With the analytic structure at hand, one can write down dispersion relations. Dispersion relations serve as a bridge between UV physics, which is hard to access, and IR physics, which we have access to. Features of a physical theory in the UV, such as unitarity and causality, often leave imprints on the IR theory, for instance, positivity bounds in EFT coefficients \cite{Adams:2006sv, Nicolis:2009qm,deRham:2017avq, deRham:2017zjm, Fuks:2020ujk, Chala:2021wpj, Li:2021cjv, deRham:2022hpx} (also see \cite{Conjecture} where these ideas are applied to the EFT of inflation). 

Inspired by the progress of the S-matrix program, there has been a lot of interest in the cosmological bootstrap program in the hope to better understand inflation \cite{Arkani-Hamed:2018kmz,Baumann:2022jpr}. At the perturbative level, there has been progress on the consequences of physical principles for the cosmological wavefunction, one of the main observable of interest. For instance, studying unitarity has led to the Cosmological Optical Theorem \cite{COT, Cespedes:2020xqq, Goodhew:2021oqg, Melville:2021lst}, which allows us to derive cutting rules for the wavefunction akin to the Cutkosky cutting rules. Studying locality led to the Manifest Locality Test for massless scalars and gravitons \cite{MLT}. These results have been useful in bootstrapping the wavefunction: for example, the form of the wavefunction for massless scalar has been bootstrapped at tree level, and there has been similar success for the graviton wavefunction as well \cite{Bonifacio:2022vwa, CSST}. However, at the end of the day, these are all perturbative results. If we want to move beyond perturbation theory, and achieve something similar to a positivity bound in amplitudes, clearly we require a non-perturbative notion of analyticity.

One viewpoint is to leverage the unitary irreducible representation of the de Sitter isometry group. One can write down the analytic structure of cosmological correlators in terms of the scaling dimension of operators present in the theory, and this idea has been studied extensively \cite{DiPietro:2021sjt,Hogervorst:2021uvp, Loparco:2023rug}. However, it is also important to note that the de Sitter isometry must be broken during inflation: in fact, any inflationary correlators with full de Sitter isometry is slow roll suppressed, meaning it is too small to be observable in the foreseeable future \cite{Green:2020ebl}. Therefore, it is also imperative to consider an alternative viewpoint that is not reliant on the isometry group of de Sitter.

This leads us to the analytic wavefunction, which incorporates many ideas from the S-matrix program \cite{Salcedo2022}. One first defines an off-shell wavefunction, where the external energies of the wavefunction are sent off-shell. One then proceeds to write down the analytic structure in terms of one of the off-shell energies. Once again, physical principles have great influence on the analytic structure of the wavefunction: by causality, the off-shell wavefunction must not containt any singularity in the lower half of the complex plane of its external energy, which is a non-perturbative result. 

For flat space, the wavefunction can be written as a canonical form of a polytope. The structure of these polytopes and their relation to physical properties has been studied extensively \cite{Arkani-Hamed:2017fdk, Arkani-Hamed:2018bjr, Benincasa:2018ssx, Benincasa:2019vqr, Benincasa:2021qcb, Benincasa:2022omn, Albayrak:2023hie}, and they obey simple recursion relations. Leveraging this simple recursion representation, the analytic structure of the off-shell wavefunction can be written down: there can only be poles and branch cut at the negative real axis of the complex plane of its external energy. Each pole and branch cut corresponds to the vanishing of energy entering a subgraph of a Feynman diagram. Armed with the analytic structure, dispersion relations for the wavefunction can be written down, which allow us to write down UV/IR sum rules. 

\begin{figure}
        \centering
    \includegraphics[scale=1.2]{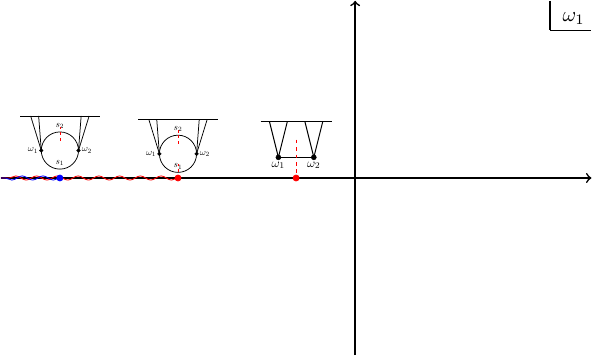}
    \caption{The analytic structure of the wavefunction in terms of external energy $\omega_1$. There are only poles and branch cuts at the negative real axis. Poles are associated with tree level diagrams, while branch cuts are associated with loop level diagrams. The singularities can also be classified into amplitude-type singularities (here they are labeled in red) and wavefunction-type singularities (here they are labeled in blue)}.
    \label{summaryfigure}
\end{figure}

On the surface, singularities for amplitudes and the flat space wavefunction look very different: for amplitudes singularities are associated with sending internal propagators on-shell (i.e. cutting internal lines), while wavefunction singularities are associated with vanishing sub-energies of diagrams (i.e. circling sub-graphs of Feynman diagram). However, upon further inspection, the locations of the singularities are seemingly related. For instance, for a one loop bubble diagram, there is a singularity for amplitudes located at $s=4m^2$, while for the wavefunction there is a singularity located at $\omega_1=-\sqrt{k^2+4m^2}$. Since $s=\omega_1^2-k^2$, these two singularities are the same! If we keep going through the list of singularities in the wavefunction, we find more and more singularities which resembles normal threshold of amplitudes.

Most of our understanding of the analytic wavefunction comes from the recursion representation of the wavefunction \cite{Arkani-Hamed:2017fdk}. While this form is simple and elegant in its own right, it looks very different from expressions commonly found for amplitudes. Therefore, it is worth asking the following question: is there a different representation that resembles Feynman integrals commonly encounted in amplitudes? If there is such a representation, it may allow us to directly import to the wavefunction ideas used in the S-matrix program, and potentially save us a lot of effort.

\begin{figure}[h!]
    \centering
    \includegraphics[scale=1.6]{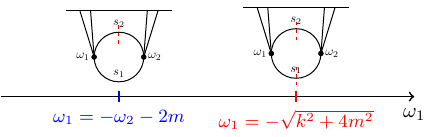}
    \caption{A summary of the singularities present for the two site one loop graph. The red singularity here is an amplitude-type singularity, while the blue singularity is a wavefunction-type singularity.}
    \label{fig:2sitesummary}
\end{figure}

\begin{figure}[h!]
    \centering
    \includegraphics{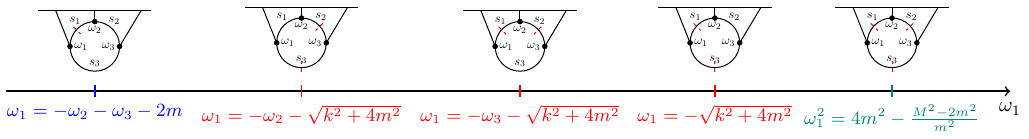}
    \caption{Singularities of the three site one loop graph. The red ones are the normal amplitude-type singularities, the teal one is the anomalous amplitude-type singularity and the blue one is the wavefunction-type singularity.}
    \label{fig:3sitesummary}
\end{figure}

In this paper we develop the \textit{amplitude representation} of the flat space wavefunction. Notably, the wavefunction can be written as an \textit{amplitude-like part} (which bears a striking resemblance to the Feynman integrals used for amplitudes) integrated against an \textit{energy-fixing kernel} (which fixes internal energies of the wavefunction). The full expression is given in \eqref{Wfnmaster}. Using this representation we can show that the wavefunction can be separated in two parts: the amplitude part (which is just the amplitude divided by thetotal energy), and a sum of sub-leading terms. These sub-leading terms are not divergent as $\omega_T$ (the total energy) goes to zero, and are less UV divergent than the amplitude. By writing the wavefunction in this way, we can classify the singularities into two sets: amplitude-type singularities associated with the amplitude part, and wavefunction-type singularities associated with the sub-leading terms. For examples of this categorization, see figure \ref{summaryfigure}, figure \ref{fig:2sitesummary} and figure \ref{fig:3sitesummary}.

The eventual hope is that this representation will allow us to directly import results from S-matrix program to the wavefunction, for instance importing positivity bounds from amplitude to the wavefunction. In addition, it has been known that cosmological wavefunctions can often be obtained from the flat space wavefunction by taking derivatives of external energies \cite{Hillman:2021bnk}.  In the amplitude representation of the wavefunction, these only affect the energy-fixing kernel, so perhaps it is possible that many of the features of the flat space wavefunction will also be present in the cosmological wavefunction, up to some slight modifications.

The outline of the paper is as follow: in Section \ref{Sec:Formalism} we derive the amplitude representation of the wavefunction. In Section \ref{sec:wfnasamp} we use the amplitude representation to write down the one loop wavefunction in terms of an amplitude and subleading terms. In Section \ref{sec:analyticity} we study the singularities of two one loop wavefunctions explicitly, and show that the singularities are indeed separated into amplitude-type singularities and wavefunction-type singularities.  In Section \ref{sec:correlators} we make some remarks on the analytic structure of in-in correlator, which is another important observable in cosmology, and how it is different from the analytic structure of the wavefunction. Finally in Section \ref{sec:outlook} we will conclude with a discussion on future directions.

\paragraph{Notations and convention}
Here the wavefunction is defined to be:
\begin{align}
    \Psi[\phi;t]=\int_{\text{BD}}^{\Phi(t)=\phi} [d\Phi] e^{iS[\Phi ; t]}\,,
\end{align}
where BD refers to Bunch-Davies vacuum in the infinite past and $S$ is the action of a theory. The wavefunction is parameterized as:
\begin{align}\label{psin}
\Psi[\phi;t]=\exp\left[   +\sum_{n}^{\infty}\frac{1}{n!} \int_{\bfk_{1},\dots\bfk_{n}}\,  \dt  \left( \sum_{a}^{n} \bfk_{a} \right) \psi_{n}(\{\bfk\};t)  \phi(\bfk_{1})\dots \phi(\bfk_{n})\right]\,,
\end{align}
Here $\{\bfk\}$ means the collection of external momentum $\{\bfk_1,\bfk_2\dots\bfk_n\}$.

We use $\Omega_k$ to label energy when it is "on-shell". If the particle has mass $m$, it is given by:
\begin{equation}
    \Omega_k=\sqrt{k^2+m^2}.
\end{equation}
The wavefunction is a function of external energies $\Omega_k$ as well as external momenta $\bfk$:
\begin{equation}
    \psi_{n}(\{\Omega\},\{\bfk\}).
\end{equation}
We refer to this as the \textit{on-shell wavefunction}.

\begin{figure}[h!]
    \centering
    \includegraphics[scale=1.3]{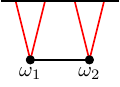}
    \caption{Tree level exchange diagram}
    \label{fig:tree_exchange}
\end{figure}

For flat space, only the total external energy entering a vertex matters if the interactions do not have derivatives. As an example, for an exchange diagram (see figure \ref{fig:tree_exchange}), even though there are four external lines, we only need to specify the energy entering the left and right vertex as well as the energy of the internal line, rather than requiring the energy of all four external lines. For this reason we will work with the \textit{off-shell wavefunction}, which is specified by considering off-shell energies $\omega_A$ (which are independent of momentum $\{\bfk\}$) on each vertex and external momenta $\bfk$. The wavefunction is written as:
\begin{equation}
    \psi_n(\{\omega\},\{\bfk\}).
\end{equation}
Here $n$ denotes the number of verticies with external legs attached. We can recover the usual on-shell wavefunction by setting $\omega_A=\sum_{a\in A}\Omega_{a}$, where $a$ runs over the external legs attached to the vertex $A$.

We will also use a slightly different set of Feynman rules from \cite{Salcedo2022} (see Appendix B of the paper). The main difference is that each vertex gains a factor of $i$, and there is no $i^{L-1}$ sitting in front. This is because the bulk-to-bulk propagator used is different by a factor of $i$ here, see \eqref{b2bpropagator}.


\section{The amplitude representation of the wavefunction}\label{Sec:Formalism}
In this section we will develop the amplitude representation of the wavefunction used in the remainder of this paper. We will show that the flat space bulk-to-bulk propagator can be written in a form reminiscent to the Feynman propagator for amplitudes. We will then use this form of the propagator to write down the wavefunction, and show that the wavefunction can be written as an amplitude-like expression integrated over an "energy-fixing kernel". We will then show how to write the simplest example of the wavefunction with one internal line, the two site exchange diagram, in this form, and show that it is equivalent to the usual recursion expression developed in \cite{Arkani-Hamed:2017fdk}. We will also make a passing remark about how this representation may relate the cosmological optical theorem to the usual optical theorem used in amplitudes.
\subsection{Bulk-to-bulk propagator}
Let's start with the usual expression for the bulk-to-bulk propagator. For flat space it is given by the following:
\begin{equation}
	G(\bfp,t_1,t_2)=\frac{1}{2\Omega_p}\left(\theta(t_1-t_2)e^{i\Omega_p(t_2-t_1)}+\theta(t_2-t_1)e^{i\Omega_p(t_1-t_2)}-e^{i\Omega_p(t_1+t_2)}\right).\label{b2bpropagator}
\end{equation}
This propagator is simply the Feynman propagator plus a homogeneous piece, which enforces the boundary condition that the bulk-to-bulk propagator vanishes as $t_1 \text{or} t_2\rightarrow 0$.

Now we can use the following:
\begin{equation}
	\theta(t_1-t_2)=\int_{-\infty}^{\infty}\frac{ds}{2\pi i}\frac{-e^{is(t_2-t_1)}}{s+i\epsilon},
\end{equation}
Unlike the usual case with the Feynman propagator, we also have this extra boundary piece $e^{i\Omega_p (t_1+t_2)}$, and we must also convert the boundary piece into an integral form as well. We then obtain:
\begin{equation}
		G(\bfp,t_1,t_2)=\int_{-\infty}^{\infty}\frac{ds}{2\pi i}\frac{-e^{is(t_1-t_2)}+e^{is(t_1+t_2)}}{s^2-\Omega_p^2+i\epsilon}.\label{Gsimple}
\end{equation}
Here this $s$ integral is understood as a contour integral which closes in the lower half complex plane. The first term is simply the Feynman propagator, and the second term is the boundary piece.
From now on I will suppress the integration region for brevity. In addition, I will also sometimes refer to $s$ as internal energy.

Now we symmetrize with respect to $s$, and we obtain the following expression:
\begin{equation}
	G(\bfp,t_1,t_2)=\frac{1}{2}\int\frac{ds}{2\pi i}\frac{(e^{ist_1}-e^{-ist_1})(e^{ist_2}-e^{-ist_2})}{s^2-\Omega_p^2+i\epsilon}.
\end{equation}

Notice how this looks like a Feynman propagator dressed with exponential factors. This is what allow us to write the wavefunction in a form similar to an amplitude: take this expression for the bulk-to-bulk propagator, substitute this into the wavefunction, then carry out the time integral first. 

Similar representations have been derived in \cite{Meltzer:2021zin, Bittermann:2022nfh}, where it is used to study the analyticity of the wavefunction at tree level. We shall see that this representation can be used to study the analyticity of the wavefunction at loop level as well.

\paragraph{Writing down the wavefunction}
Given a Feynman diagram, we can write down the flat space wavefunction as:
\begin{equation}
    \psi_n=\int\prod_i \left[\frac{ds_1}{2\pi i}\dots\frac{ds_I}{2\pi i}\right]\prod_{a}\left[\int_{-\infty}^{0}dt_a \,i e^{i\omega_a t_a}\right]\int\prod_l \frac{d^D p_l}{(2\pi)^D}\prod_i \frac{(e^{is_it_1}-e^{-is_it_1})(e^{is_it_2}-e^{-is_it_2})}{s_i^2-\Omega_{p_i}^2+i\epsilon},\label{General}
\end{equation}
Here $i$ labels bulk-to-bulk propagators (and $I$ is the total number of bulk-to-bulk propagators), $a$ labels each vertex, $l$ labels the loop momentum to be integrated over, and $t_l, t_r$ labels to which vertex the propagator is attached. If we take this expression, integrate over $s$ and then integrate over $t$, we would obtain the usual recursion expression. 

The time integrals on each individual vertex factorize, and are straightforward to do. For a vertex with $m$ internal propagators attached, it has the form:
\begin{equation}
    \tilde{D}_a=i\int_{-\infty}^{0}dt_a\,e^{i\omega_a t_a}\prod_{j=1}^{m}(e^{i s_j t_a}-e^{-i s_j t_a})=\int_{-\infty}^{0}dt_a\,e^{i\omega_a t_a}\prod_{j=1}^{m}\left(\sum_{\sigma_j=\pm}\sigma_j e^{i \sigma_j s_j t_a}\right).
\end{equation}
We also regulate the integral by sending $\omega\rightarrow\omega-i\epsilon$. This gives:
\begin{equation}
    \tilde{D}_a=\sum_{\sigma_j=\pm}\frac{\sigma_1\sigma_2\dots\sigma_m}{\omega_a+\sum_j\sigma_js_j-i\epsilon}. \label{tildDa}
\end{equation}
Let us go through a simple example where there are two bulk-to-bulk propagators attached to vertex $a$. The time integral has the form:
\begin{equation}
    \tilde{D}_a=i\int_{\infty}^{0}dt_a\,e^{i\omega_a t_a}(e^{i s_1 t_a}-e^{-i s_1 t_a})(e^{i s_2 t_a}-e^{-i s_2 t_a}).
\end{equation}
It is not hard to see that this gives:
\begin{equation}
    \tilde{D}_a=\frac{1}{\omega_a+s_1+s_2-i\epsilon}-\frac{1}{\omega_a-s_1+s_2-i\epsilon}-\frac{1}{\omega_a+s_1-s_2-i\epsilon}+\frac{1}{\omega_a-s_1-s_2-i\epsilon}, \label{da2}
\end{equation}
which is indeed the \eqref{tildDa} for $j=2$.

Putting this back into the expression \eqref{General} gives us:
\begin{equation}
	\psi_n=\frac{1}{2^I}\int\left[\frac{ds_1}{2\pi i}\dots\frac{ds_I}{2\pi i}\right]\prod_a\tilde{D}_a(\omega_a,\{s\})\int\prod_l \frac{d^D p_l}{(2\pi)^D}\prod_i \frac{1}{s_i^2-\Omega_i^2+i\epsilon}. \label{Wfnmaster}
\end{equation} 

This representation for the wavefunction holds for graphs with any topology at any loop order. Notice how the momentum integral resembles the amplitude of the same Feynman diagram in $D$-dimension, which is given by:
\begin{equation}
    \mathcal{A}_n=
    \int\prod_l\frac{d^D p_l}{(2\pi)^D}\prod_i \frac{1}{\Omega_i^2-i\epsilon}.
\end{equation}
I will call the momentum integral the amplitude-like part of the wavefunction. Also, when integrating over $s$, $\prod_{a}\tilde{D}_a$ fixes $s$ in terms of $\omega$, the external energies. For this reason I will call it the the energy-fixing kernel (or just kernel in short).

Readers familiar with the wavefunction may wonder if this is connected with the fact that the total energy pole of the wavefunction is the amplitude of the same diagram. We shall see in the next section that this is indeed the case: when we expand the energy-fixing kernel and do the $s$-integral, we will obtain the amplitude for a $d+1$-dimension amplitude alongside some subleading terms.

\subsection{First example: tree level exchange}
In order to demonstrate how the formalism works, let us consider a two site tree level exchange diagram. 

The wavefunction coefficient is:
\begin{equation}
	\psi_2=\frac{(-i)^2}{2}\int^{0}_{-\infty} dt_1\int^{0}_{-\infty} dt_2\int\frac{ds}{2\pi i}e^{i\omega_1t_1}\frac{(e^{ist_1}-e^{-ist_1})(e^{ist_2}-e^{-ist_2})}{s^2-\Omega_p^2+i\epsilon}e^{i\omega_2t_2}
\end{equation}
Now do the time integral:
\begin{equation}
	\psi_2=\frac{1}{2}\int\frac{ds}{2\pi i}\left[\frac{1}{\omega_1+s-i\epsilon}-\frac{1}{\omega_1-s-i\epsilon}\right]\left[\frac{1}{\omega_2+s-i\epsilon}-\frac{1}{\omega_2-s-i\epsilon}\right]\frac{1}{s^2-\Omega_p^2+i\epsilon}
\end{equation}
This can be simplified into:
\begin{equation}
	\psi_2=\frac{1}{2}\int\frac{ds}{2\pi i}\frac{4s^2}{(\omega_1^2-s^2-i\epsilon)(\omega_2^2-s^2-i\epsilon)}\frac{1}{s^2-p^2-m^2+i\epsilon}
\end{equation}
Here $\frac{1}{s^2-p^2-m^2+i\epsilon}$ is indeed the usual expression for the exchange diagram in amplitudes. The energy-fixing kernel is given by $\frac{4s^2}{(\omega_1^2-s^2-i\epsilon)(\omega_2^2-s^2-i\epsilon)}$. 

Let us try and recover the usual recursion expression. We need to remember that the $s$ integral contour encloses the lower half complex plane, so poles  like $s=-\omega_1+i\epsilon$ are not picked up. This gives us the isolated poles:
\begin{equation}
	\psi_2=\frac{1}{2}\left(\frac{2\omega_1}{(\omega_2^2-\omega_1^2)(\omega_1^2-\Omega_p^2)}+\frac{2\omega_2}{(\omega_1^2-\omega_2^2)(\omega_2^2-\Omega_p^2)}-\frac{2\Omega_p}{(\omega_1^2-\Omega_p^2)(\omega_2^2-\Omega_p^2)}\right).
\end{equation}
Notice this expression can also be written as:
\begin{equation}
    \psi_2=\frac{1}{\omega_1+\omega_2}\left(\frac{-\omega_1}{(\omega_1-\omega_2)(\omega_1^2-\Omega_p^2)}+\frac{\omega_2}{(\omega_1-\omega_2)(\omega_2^2-\Omega_p^2)}\right)-\frac{\Omega_p}{(\omega_1^2-\Omega_p^2)(\omega_2^2-\Omega_p^2)}.
\end{equation}
The total energy pole has been isolated from the rest of the contribution, which only has poles when $\omega_1=\pm \Omega_p$ or $\omega_2=\pm \Omega_p$.There is no folded singularity, i.e. no divergences as $\omega_1\rightarrow\omega_2$, since the terms in the bracket cancel. If we take $\omega_1+\omega_2\rightarrow 0$ we obtain:
\begin{equation}
    \underset{\omega_1+\omega_2\rightarrow 0}{\text{lim}}\psi_2=\frac{1}{\omega_1+\omega_2}\left(\frac{1}{-\omega_1^2+\Omega_p^2}\right)=\frac{1}{\omega_1+\omega_2}\left(\frac{1}{-\omega_1^2+|\bfp|^2+m^2}\right)
\end{equation}
We indeed recover the expression for the amplitude as a total energy pole.

We can further simplify this to obtain the usual recursion expression:
\begin{equation}
	\psi_2=\frac{1}{(\omega_1+\omega_2)(\omega_1+\Omega_p)(\omega_2+\Omega_p)}.
\end{equation}
In this simple example we can see how the wavefunction can be separated into a total energy pole (and its residue is the amplitude), and an extra piece which is finite when total energy is zero. We will see how this formalism allow us to do the same for a loop level wavefunction as well.

\subsection{From the Cosmological optical theorem to the optical theorem for amplitude}
Before we move on, there is an interesting feature about this new representation: namely, it relates the disc operator defined in the Cosmological Optical Theorem (COT) to the imaginary part of amplitudes.

The disc operator used in the COT is defined as follows \cite{COT}:
\begin{multline}
    \underset{k_1,\dots,k_n}{\text{Disc}}f(k_1,\dots,k_n,k_{n+1},\dots,k_{m},\{\bfk\})=f(k_1,\dots,k_n,k_{n+1},\dots,k_{m},\{\bfk\})\\-f^\ast(k_1,\dots,k_n,-k_{n+1}^\ast,\dots,-k_{m}^\ast,-\{\bfk\}).
\end{multline}

Now notice the following: the kernel is written as a product of $\tilde{D}_a$, and in general the following is true:
\begin{itemize}
    \item If the number of internal lines attached to the vertex is an odd number, we have: 
    \begin{equation}
        \underset{\{s\}}{\text{Disc}}\tilde{D}_a=0.
    \end{equation}
    \item If the number of internal lines attached to the vertex is an even number, we have:
    \begin{equation}
        \underset{\{s\}}{\text{Disc}}\tilde{D}_a=2\tilde{D}_a.
    \end{equation}
\end{itemize}

To see this, consider a vertex connected to a single bulk-to-bulk propagator. Then we have:
\begin{equation}
    \underset{s}{\text{Disc}}\left[\frac{1}{\omega+s-i\epsilon}-\frac{1}{\omega-s-i\epsilon}\right]=\frac{1}{\omega+s-i\epsilon}-\frac{1}{\omega-s-i\epsilon}-\frac{1}{-\omega+s+i\epsilon}+\frac{1}{-\omega-s+i\epsilon}=0.
\end{equation} 
Another example a vertex connected with two bulk-to-bulk propagators. Here we have:
\begin{multline}
    \underset{s}{\text{Disc}}\left[\frac{1}{\omega+s_1+s_2-i\epsilon}+\frac{1}{\omega-s_1-s_2-i\epsilon}\right]\\=\frac{1}{\omega+s_1+s_2-i\epsilon}+\frac{1}{\omega-s_1-s_2-i\epsilon}-\frac{1}{-\omega+s_1+s_2+i\epsilon}-\frac{1}{-\omega-s_1-s_2+i\epsilon}\\=2\left[\frac{1}{\omega+s_1+s_2-i\epsilon}-\frac{1}{\omega-s_1-s_2-i\epsilon}\right].
\end{multline}
Hence,
\begin{multline}
    \underset{s_1,s_2}{\text{Disc}}\left[\frac{1}{\omega+s_1+s_2-i\epsilon}-\frac{1}{\omega-s_1+s_2-i\epsilon}-\frac{1}{\omega+s_1-s_2-i\epsilon}+\frac{1}{\omega-s_1-s_2-i\epsilon}\right]\\=2\left[\frac{1}{\omega+s_1+s_2-i\epsilon}-\frac{1}{\omega-s_1+s_2-i\epsilon}-\frac{1}{\omega+s_1-s_2-i\epsilon}+\frac{1}{\omega-s_1-s_2-i\epsilon}\right].
\end{multline}
It is easy to see how this generalizes to cases with more bulk-to-bulk propagators attached.

When the disc operator acts on the energy-fixing kernel, it either vanishes or reproduce the energy-fixing kernel. It can also act on the amplitude like part, which will give us the imaginary part of the amplitude. We know in amplitudes that the imaginary part of the amplitude is related to cutting internal lines \cite{Schwartz:2014sze}. As a result this may help us link between the optical theorem in amplitudes and COT.

As an example let us consider the tree level exchange diagram. Here the disc of the energy-fixing kernel vanishes. We have (after restoring the coupling constant $\lambda$ on each vertex):
\begin{equation}
    \underset{\Omega_p}{\text{Disc }}\psi_2=\frac{1}{2}\int_{-\infty}^{\infty}\frac{ds}{2\pi i}\frac{4s^2}{(\omega_1^2-s^2-i\epsilon)(\omega_2^2-s^2-i\epsilon)}\lambda^2(-2\pi i\delta(s^2-\Omega_p^2+i\epsilon))
\end{equation}
We can use the delta function to do the integral, which gives:
\begin{equation}
    \underset{\Omega_p}{\text{Disc }}\psi_2=\frac{2\Omega_p \lambda^2}{(\omega_1^2-\Omega_p^2)(\omega_2^2-\Omega_p^2)}=\frac{\lambda^2}{2\Omega_p}\underset{\Omega_p}{\text{Disc }}\left[\frac{1}{\omega_1+\Omega_p-i\epsilon}\right]\underset{\Omega_p}{\text{Disc }}\left[\frac{1}{\omega_2+\Omega_p-i\epsilon}\right].
\end{equation}
Of course we recover the usual result from the COT. However, this may become helpful in establishing positivity bounds. For instance, since the imaginary part of the amplitude needs to have a positive residue by unitarity, we have $\lambda^2$ positive as well (i.e. $\lambda$ is real). Therefore we already can establish that $\text{Disc}\psi_2$ must be positive if $\omega_1>\Omega_p$ and $\omega_2>\Omega_p$, i.e. Disc of $\psi_2$ is positive in the physical regime.

For one loop diagrams there are a few complications: namely, the disc of the energy-fixing kernel may not vanish (for examples see Appendix \ref{app:a}). But in some cases such as the two site loop (i.e. figure \ref{fig:2siteclean}) we can still write the disc of the wavefunction as the imaginary part of the amplitude integrated against the energy-fixing kernel. We leave a systematic study of unitarity and positivity bounds for the wavefunction at loop level for the future.


\section{One loop wavefunction as amplitudes}\label{sec:wfnasamp}
This new representation of the wavefunction becomes a lot more useful when we start tackling loops. It is known that for tree level, the wavefunction has a total energy pole, and the residue gives the flat space amplitude \cite{Maldacena:2011nz, Raju:2012zr,Arkani-Hamed:2018bjr, Benincasa:2018ssx}. Using this new representation of the wavefunction, not only can we show this explicitly, we can also provide a way to obtain the remaining terms outside of the total energy pole.

In this section I will provide a heuristic argument on why this is true for one loop wavefunctions. I will show that the wavefunction can be written as an amplitude divided by the total energy, plus some remainder terms \footnote{In \cite{Benincasa:2018ssx} a similar representation was written down at tree level from the perspective of the cosmological polytope. One can then extend the representation to loop level by doing contour integrals over tree level wavefunctions. For our representation we make no explicit reference to tree level wavefunctions, however it would be interesting to see if similar relations exist in our representation as well.}. Since the wavefunction has a momentum integral, there are potentially UV divergences. I will show that the amplitude part of the wavefunction is the most UV divergent part, and every remainder term cannot contribute to UV divergences that are more severe than the amplitude. I will then write down the two site and three site one loop wavefunction explicitly in this form.
\subsection{Total energy pole}
Let us first look at a heuristic argument on why the total energy pole gives the amplitude at one loop. 

\begin{figure}[h!]
    \centering
    \includegraphics{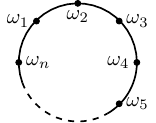}
    \caption{Diagram for an $n$ site one loop wavefunction. External lines have been omitted.}
    \label{fig:nloopclean}
\end{figure}

An $n$ site one loop wavefunction (see figure \ref{fig:nloopclean}) has the following form:
\begin{equation}
    \psi_n=\frac{1}{2^n}\int \prod_{j=1}^{n}\frac{ds_j}{2\pi i}\tilde{D}_j\int_{\bfp}\prod_{l=1}^{n}\frac{1}{s_l^2-\Omega_{p_l}^2}.
\end{equation}

Here $\tilde{D}_j$ is given by:
\begin{equation}
    \tilde{D}_j=\frac{1}{\omega_j+s_{j-1}+s_j-i\epsilon}-\frac{1}{\omega_j-s_{j-1}+s_j-i\epsilon}-\frac{1}{\omega_j+s_{j-1}-s_j-i\epsilon}+\frac{1}{\omega_j-s_{j-1}-s_j-i\epsilon}.
\end{equation}
(and I have set $s_0=s_n$).

If we expand the product of $\tilde{D}_j$ we find that the wavefunction has the form:
\begin{equation}
    \psi_n=\frac{1}{2^n}\int \prod_{j=1}^{n}\frac{ds_j}{2\pi i}\frac{1}{\omega_j+s_j-s_{j-1}-i\epsilon}\int_{\bfp}\prod_{l=1}^{n}\frac{1}{s_l^2-\Omega_{pl}^2}+(\text{Remainder})
\end{equation}
We will say more about the remainder terms below.

\begin{figure}
    \centering
    \includegraphics{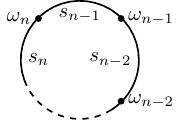}
    \caption{$n$ site one loop diagram. Consider energy flowing clockwise, and let $\omega_l$ be energy entering the loop for vertex $l$. Clearly energy is conserved at each vertex if  $s_{n-1}=\omega_n+s_n$, $s_{n-2}=\omega_{n-1}+\omega_n+s_n$, and so on.}
    \label{fig:nloop}
\end{figure}

Now do the $s_{n-1}$ contour integral. We close the contour in the lower half plane. This picks up the two poles $s_{n-1}=\omega_n+s_n$ and $s_{n-1}=\Omega_{p(n-1)}$ separately, and gives:
\begin{multline}
    \psi_n=\frac{1}{2^n}\int \frac{ds_n}{2\pi i}\left[\prod_{j=1}^{n-2}\frac{ds_j}{2\pi i}\frac{1}{\omega_j+s_j-s_{j-1}}\right]\frac{1}{\omega_{n-1}+\omega_n+s_{n}-s_{n-2}}\int_{\bfp}\prod_{l=1}^{n-2}\frac{1}{s_l^2-\Omega_{pl}^2}\frac{1}{(s_n+\omega_n)^2-\Omega_{p(n-1)}^2}\frac{1}{s_n^2-\Omega_{pn}^2}\\
    +\frac{1}{2^n}\int \frac{ds_n}{2\pi i}\left[\prod_{j=1}^{n-2}\frac{ds_j}{2\pi i}\frac{1}{\omega_j+s_j-s_{j-1}}\right]\int_{\bfp}\frac{1}{\omega_{n-1}+\Omega_{p(n-1)}-s_{n-2}}\frac{1}{\omega_n+s_n-\Omega_{p(n-1)}}\prod_{l=1}^{n-1}\frac{1}{2\Omega_{p(n-1)}}\frac{1}{s_l^2-\Omega_{pl}^2}\\
    +(\text{Remainder}).\label{npt2}
\end{multline}
We can then keep doing the contour integrals successively, i.e. do the $s_{n-2}$ integral, then the $s_{n-3}$ integral, and so on. Now notice the following:
\begin{itemize}
    \item If we keep picking up the poles only from the kernel, i.e. from the $\omega_j+s_j-s_{j-1}$, then we will get an overall $\omega_T=\sum_{j=1}^{n}\omega_j$ in front. This is because the set of $s_j-s_{j-1}$ are not linearly independent: when we do the $s_1$ integral, we would get $s_1=\sum_{j=2}^{n}\omega_j+s_n$, and substituting this into $\omega_1+s_1-s_n$ gives the total energy pole.
    \item If we pick up any poles from the propagators, i.e. take $s_j=\Omega_{pj}$, then the $\Omega_{pj}$ enter into the kernel, and we would not get a total energy pole at the end.
\end{itemize}
Let's look at the total energy pole first. The total energy pole is:
\begin{equation}
    \psi_n=\frac{1}{2^n}\frac{1}{\omega_T}\int\frac{ds_n}{2\pi i}\int_{\bfp}\prod_{l=1}^{n}\frac{1}{\tilde{s}_l^2-p_n^2-m^2}+\dots.
\end{equation}
Here $\tilde{s}_l=s_n+\sum_{j=l+1}^{n}\omega_j$. This is just integrand for an amplitude: energy at each vertex is conserved (see the diagram \ref{fig:nloop}), and we obtain the $\int d^{d+1}p$ integral measure for the amplitude by setting $s_n=p_0$. Also worth noting is that the whole integral scales as $p^{D+1-2n}$ in $D$-dimensions.

What about the other poles? As an example let us consider the second line of \eqref{npt2}. If we do the other integrals and take the poles from the kernel this gives\footnote{One needs to be careful about the $i\epsilon$ prescription here: For example, take $s=p-i\epsilon_p$, I may get terms like $\omega_j+s_j-p+i\epsilon_p-i\epsilon_j$. Depending on the size of $\epsilon_p$ and $\epsilon_j$ this pole may not be picked up.}:
\begin{equation}
    \int_{\bfp}\frac{1}{2\Omega_{p(n-1)}(\omega_T-\omega_{n-1}-\Omega_{p(n-1)})}\prod_{l=1}^{n-1}\frac{1}{s_l^2-\Omega_{pl}^2}
\end{equation}
This integral scales as $p^{D-2n}$ in $D$-dimensions. It is less divergent than the total energy pole. A similar story applies to any other terms which we pick up a pole from the propagators.

\paragraph{The "remainder" terms} Let's talk about the remainder terms in \eqref{npt2}. In general, each term looks like:
\begin{equation}
    \int \prod_{j=1}^{n}\frac{ds_j}{2\pi i}\frac{1}{\omega_j \pm s_j\pm s_{j-1}-i\epsilon}\int_{\bfp}\prod_{l=1}^{n}\frac{1}{s_l^2-\Omega_{pl}^2}
\end{equation}
\begin{itemize}
    \item If the set of $\pm s_j\pm s_{j-1}$ in the kernel of the term are linearly independent, then one can always do all the $s$ integrals without picking up any poles from the propagator. In this case we would not get a total energy pole, and the resulting momentum integral scales like $p^{D-2n}$.
    \item Whenever we pick up poles from the propagator we get a momentum integral which scales like $p^{D-2n}$, for reasons explained above.
\end{itemize}
As a result the remainder terms can never lead to UV divergences that are more severe than the total energy pole. For instance, in $D=3$ and $n=2$, we expect that only the total energy pole has a UV divergence, while the remainder terms are all finite. We will confirm this soon. 

\subsection{Example: two site loop}
Let us look at the simple example of a two site loop. Here we will write down the wavefunction in terms of the amplitude $\mathcal{A}_2$, as well as $\psi_2^{\text{sub}}$, a collection of terms which are subleading as $\omega_T\rightarrow 0$ and are less UV divergent than the amplitude. 

\begin{figure}[h!]
    \centering
    \includegraphics[scale=1.3]{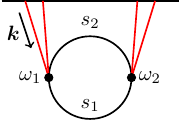}
    \caption{Two site one loop diagram}
    \label{fig:2siteclean}
\end{figure}

The wavefunction is:
\begin{equation}
    \psi_2=\frac{1}{4}\int\frac{ds_1}{2\pi i}\frac{ds_2}{2\pi i} \tilde{D}_1\tilde{D}_2\int_{\bfp}\frac{1}{s_1^2-\Omega_{p1}^2}\frac{1}{s_2^2-\Omega_{p2}^2}, \label{ARpsi2}
\end{equation}
where $\tilde{D}_1$ is given by:
\begin{equation}
    \tilde{D}_1=\frac{1}{\omega_1+s_1+s_2-i\epsilon}-\frac{1}{\omega_1-s_1+s_2-i\epsilon}-\frac{1}{\omega_1+s_1-s_2-i\epsilon}+\frac{1}{\omega_1-s_1-s_2-i\epsilon},\label{tilD1}
\end{equation}
and $\tilde{D}_2$ is obtained simply by replacing $\omega_1$ with $\omega_2$. We can simplify this expression\footnote{Or just use the expression \eqref{Gsimple} and do the time integral} to obtain:
\begin{multline}
   \psi_2=\int\frac{ds_1}{2\pi i}\frac{ds_2}{2\pi i}\frac{1}{\omega_1+s_1+s_2-i\epsilon_1}\left[\frac{1}{\omega_2+s_1+s_2-i\epsilon_2}-\frac{1}{\omega_2-s_1+s_2-i\epsilon_2}\right.\\\left.-\frac{1}{\omega_2+s_1-s_2-i\epsilon_2}+\frac{1}{\omega_2-s_1-s_2-i\epsilon_2}\right]\int_{\bfp}\frac{1}{s_1^2-\Omega_{p1}^2+i\epsilon_{p1}}\frac{1}{s_2^2-\Omega_{p2}^2+i\epsilon_{p2}}. \label{psi2integral}
\end{multline}
To simplify the wavefunction, do the $s_2$ integral first, where the contour is closed in the lower half plane\footnote{I will also take $\epsilon_2<\epsilon_{p2}$}. The total energy pole comes from the fourth term in the bracket in \eqref{psi2integral}, which integrates to:
\begin{equation}
    \psi_2=\int\frac{ds_1}{2\pi i}\int_{\bfp}\frac{-1}{\omega_T}\frac{1}{(s_1^2-\Omega_{p1}^2)((\omega_2-s_1)^2-\Omega_{p2}^2)}+\frac{1}{(\omega_1+s_1+p_2)(\omega_2-s_1-p_2)}\frac{1}{2\Omega_{p2}(s_1^2-\Omega_{p1}^2)}+\dots
\end{equation}
Relabelling $s_1$ as $p_0$, we have:
\begin{equation}
    \psi_2=\frac{i}{\omega_T}\int\frac{d^{D+1}p}{(2\pi)^{D+1}}\frac{1}{(p_0^2-|\bfp|^2-m^2)((\omega_2-p_0)^2-|\bfk-\bfp|^2-m^2)}+\psi_2^{\text{sub}}
\end{equation}
The integrand is exactly the amplitude, where the external four momentum entering the loop is $(\omega_2,\bfk)$. However since the result should be symmetric with respect to exchange of $\omega_1$ and $\omega_2$, we symmetrize the result to obtain:
\begin{equation}
    \psi_2=\frac{1}{2\omega_T}\left(\mathcal{A}(-\omega_1^2+|\bfk|^2)+\mathcal{A}(-\omega_2^2+|\bfk|^2)\right)+\psi_2^{\text{sub}}, \label{psi2}
\end{equation}
where 
\begin{equation}
    \mathcal{A}(-\omega^2+|\bfk|^2)=i\int\frac{d^{d+1}p}{(2\pi)^{d+1}}\frac{1}{(p_0^2-|\bfp|^2-m^2)((\omega-p_0)^2-|\bfk-\bfp|^2-m^2)}
\end{equation}

\paragraph{Subleading terms} 
In this simple case we can write down $\psi_2^{\text{sub}}$ explicitly. It is given by:
\begin{align}
    \psi_2^{\text{sub}}&=\psi_2^{\text{FF}}+\psi_2^{\text{FB}}+\psi_2^{\text{BB}},\\
    \psi_2^{\text{FF}}&=\int\frac{ds_1}{2\pi i}\int_{\bfp}\frac{1}{(\omega_1+s_1+\Omega_{p2})(\omega_2-s_1-\Omega_{p2})}\frac{1}{2\Omega_{p2}(s_1^2-\Omega_{p1}^2)},\\
    \psi_2^{\text{FB}}&=-\int\frac{ds_1}{2\pi i}\int_{\bfp}\frac{1}{(\omega_1+s_1+\Omega_{p2})(\omega_2-s_1+\Omega_{p2})}\frac{1}{2\Omega_{p2}(s_1^2-\Omega_{p1}^2)}-(\Omega_{p1}\leftrightarrow \Omega_{p2}),\\
    \psi_2^{\text{BB}}&=\int_{\bfp}\frac{1}{(\omega_1+\Omega_{p1}+\Omega_{p2})(\omega_2+\Omega_{p1}+\Omega_{p2})}.
\end{align}
There is a way to understand why the remaining terms are organised in this way. Notice that:
\begin{equation}
		G(\bfp,t_1,t_2)=\int_{-\infty}^{\infty}\frac{ds}{2\pi i}\frac{e^{is(t_1+t_2)}-e^{is(t_1-t_2)}}{s^2-\Omega_p^2+i\epsilon}=G_{F}(\bfp,t_1,t_2)-G_{B}(\bfp,t_1,t_2),
\end{equation}
where $G_F(\bfp,t_1,t_2)$ is the Feynman propagator and $G_{B}(\bfp,t_1,t_2)$ is the boundary term added. Therefore, we can always organise $\psi_2$ as:
\begin{equation}
    \psi_2=\int dt_1 \int dt_2\int_{\bfp}e^{i \omega_1 t_1}e^{i \omega_2 t_2}\left(G_F(p_1)G_F(p_2)-G_F(p_1)G_B(p_2)-G_B(p_1)G_F(p_2)+G_B(p_1)G_B(p_2)\right).
\end{equation}
Now it is clear why we can separate the terms as written above:
\begin{itemize}
    \item The total energy pole, i.e. $\mathcal{A}/\omega_T$, always comes from the $G_FG_F$ term, since the amplitude is computed with Feynman propagators in the first place. However, we get extra terms that are encapsulated in $\psi_2^{\text{FF}}$, coming from picking up the pole in the propagator in the $s_2$ integral.
    This reflects the fact that we are not integrating time from $-\infty$ to $\infty$, and so there are extra terms to compensate.
    \item The $\psi_2^{\text{BB}}$ term comes from the $G_BG_B$ term. No nested time integrals are required to evaluate this term, hence its form is simpler than the other remainder term.
    \item The $\psi_2^{\text{FB}}$ term comes from the $G_BG_F$ term, i.e. we are mixing contributions from the Feynman propagator and the boundary term. In this simple case $\psi_2^{\text{FB}}$ can actually be further simplified to be:
    \begin{multline}
        \psi_2^{\text{FB}}=-\int_{\bfp}\frac{1}{4\Omega_{p1}\Omega_{p2}}\left[\frac{1}{(\omega_1+\Omega_{p1}+\Omega_{p2})(\omega_1+\omega_2+2\Omega_{p1})}+\frac{1}{(\omega_2+\Omega_{p1}+\Omega_{p2})(\omega_1+\omega_2+2\Omega_{p1})}\right.\\\left. +\frac{1}{(\omega_1+\Omega_{p1}+\Omega_{p2})(\omega_1+\omega_2+2\Omega_{p2})}+\frac{1}{(\omega_2+\Omega_{p1}+\Omega_{p2})(\omega_1+\omega_2+2\Omega_{p2})}\right].\label{FBpsi2}
    \end{multline}
    As we will see in the next section, these terms will give rise to singularities that are not present in amplitudes. 
\end{itemize}
\subsection{Example: three site loop}
\begin{figure}[h!]
    \centering
    \includegraphics[scale=1.5]{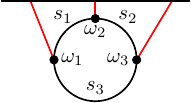}
    \caption{The three site one loop graph.}
    \label{fig:3siteclean}
\end{figure}

Let us move on to the example of a three site loop (see figure \ref{fig:3siteclean}). 

The wavefunction is:
\begin{equation}
    \psi_3=\frac{1}{8}\int\frac{ds_1}{2\pi i}\frac{ds_2}{2\pi i}\frac{ds_3}{2\pi i}\tilde{D}_1\tilde{D}_2\tilde{D}_3\int_{\bfp}\frac{1}{s_1^2-\Omega_{p1}^2}\frac{1}{s_2^2-\Omega_{p2}^2}\frac{1}{s_3^2-\Omega_{p3}^2}.
\end{equation}
Here we have:
\begin{equation}
    \tilde{D}_i=\frac{1}{\omega_i+s_i+s_{i-1}-i\epsilon}-\frac{1}{\omega_i-s_i+s_{i-1}-i\epsilon}-\frac{1}{\omega_i+s_i-s_{i-1}-i\epsilon}+\frac{1}{\omega_i-s_i-s_{i-1}-i\epsilon},
\end{equation}
and $s_0=s_3$. Once again we can simplify this to (suppressing the $i\epsilon$ in the denominator of the kernel):
\begin{multline}
    \psi_3=\int\frac{ds_1}{2\pi i}\frac{ds_2}{2\pi i}\frac{ds_3}{2\pi i}\left(\frac{1}{k_1+s_1-s_3}-\frac{1}{k_1+s_1+s_3}\right)\left(\frac{1}{k_2+s_2-s_1}-\frac{1}{k_2+s_2+s_1}\right)\\\left(\frac{1}{k_3+s_3-s_2}-\frac{1}{k_3+s_3+s_2}\right)\int_{\bfp}\frac{1}{s_1^2-\Omega_{p1}^2}\frac{1}{s_2^2-\Omega_{p2}^2}\frac{1}{s_3^2-\Omega_{p3}^2}.
\end{multline}
Taking into account that only poles in the lower half $s$ planes are picked up, we find that the term which contributes to the amplitude is:
\begin{equation}
    \psi_3=\int\frac{ds_1}{2\pi i}\frac{ds_2}{2\pi i}\frac{ds_3}{2\pi i}\frac{1}{k_1+s_1-s_3}\frac{1}{k_2+s_2-s_1}\frac{1}{k_3+s_3-s_2}\int_{\bfp}\frac{1}{s_1^2-\Omega_{p1}^2}\frac{1}{s_2^2-\Omega_{p2}^2}\frac{1}{s_3^2-\Omega_{p3}^2}+\dots.
\end{equation}
Performing the $s_1$ and $s_2$ integrals give:
\begin{equation}
    \psi_3=\int\frac{ds_3}{2\pi i}\int_{\bfp}\frac{1}{\omega_T}\frac{1}{(\omega_2+\omega_3+s_3)^2-\Omega_{p1}^2}\frac{1}{(\omega_3+s_3)^2-\Omega_{p2}^2}\frac{1}{s_3^2-\Omega_{p3}^2}+\psi_3^{\text{sub}}
\end{equation}
Once again the integral here is simply the integral for the amplitude once we replace $s_3\rightarrow p_0$. Therefore we have:
\begin{equation}
    \psi_3=\frac{1}{\omega_T}\mathcal{A}(-\omega_2^2+k_2^2,-\omega_3^2+k_3^2)+\psi_3^{\text{sub}}.
\end{equation}
Similar to the two site case, we can organise the remainder terms as:
\begin{equation}
    \psi_3^{\text{sub}}=\psi_3^{\text{FFF}}+\psi_3^{\text{FFB}}+\psi_3^{\text{FBB}}+\psi_3^{\text{BBB}}.
\end{equation}
Writing out the remainder terms explicitly:
\begin{multline}
    \psi_3^{\text{FFF}}=\int\frac{ds_3}{2\pi i}\int_{\bfp}\frac{1}{2\Omega_{p2}}\frac{1}{\omega_1+\omega_2+\Omega_{p2}-s_3}\frac{1}{\omega_3-\Omega_{p2}+s_3}\frac{1}{s_3^2-\Omega_{p3}^2}\\+\frac{1}{2\Omega_{p1}}\frac{1}{\omega_2+\omega_3-\Omega_{p1}+s_3}\frac{1}{\omega_1+\Omega_{p1}-s_3}\frac{1}{(\omega_3+s_3)^2-\Omega_{p2}^2}\frac{1}{s_3^2-\Omega_{p3}^2}\\+\frac{1}{4\Omega_{p1}\Omega_{p2}}\frac{1}{\omega_2-\Omega_{p1}+\Omega_{p2}}\frac{1}{\omega_1+\Omega_{p1}-s_3}\frac{1}{\omega_3-\Omega_{p2}+s_3}\frac{1}{s_3-\Omega_{p3}^2}.
\end{multline}
\begin{multline}
    \psi_3^{\text{FFB}}=\int\frac{ds_2}{2\pi i}\frac{ds_3}{2\pi i}\int_{\bfp}\frac{1}{\omega_2+\Omega_{p1}+s_2}\frac{1}{\omega_1+\Omega_{p1}-s_3}\frac{1}{\omega_3-s_2+s_3}\frac{1}{2\Omega_{p1}}\frac{1}{s_2^2-\Omega_{p2}^2}\frac{1}{s_3^2-\Omega_{p3}^2}+(\text{permutations}),\label{psi3FFB}
\end{multline}
\begin{multline}
    \psi_3^{\text{FBB}}=\int\frac{ds_2}{2\pi i}\int_{\bfp}\frac{1}{\omega_2+\Omega_{p1}+\Omega_{p2}}\frac{1}{\omega_1+\Omega_{p1}-s_3}\frac{1}{\omega_3-\Omega_{p2}+s_3}\frac{1}{2\Omega_{p1}}\frac{1}{2\Omega_{p2}}\frac{1}{s_3^2-\Omega_{p3}^2}+(\text{permutations}),
\end{multline}
\begin{multline}
    \psi_3^{\text{FFF}}=\int_{\bfp}\frac{1}{\omega_2+\Omega_{p1}+\Omega_{p2}}\frac{1}{\omega_1+\Omega_{p1}-\Omega_{p3}}\frac{1}{\omega_3-\Omega_{p2}+\Omega_{p3}}\frac{1}{2\Omega_{p1}}\frac{1}{2\Omega_{p2}}\frac{1}{2\Omega_{p3}}+(\text{permutations}).
\end{multline}
The momentum integral for the amplitude scales as $p^{-2}$ while the remainder terms all scale as $p^{-3}$ for $D=3$.
\paragraph{Summary}
Let us summarize our observations so far.
\begin{itemize}
    \item The one loop $n$ site wavefunction can be written as:
    \begin{equation}
        \psi_n=\frac{\mathcal{A}_n}{\omega_T}+\psi_n^{\text{sub}}.\label{amprepn}
    \end{equation}
    Here $\mathcal{A}_n$ is the one loop $n$ site amplitude.
    \item The remaining parts of the wavefunction, $\psi_n^{\text{sub}}$, are subleading as $\omega_T\rightarrow 0$ and are less UV divergent than the amplitude part of the wavefunction. 
\end{itemize}

\paragraph{Generalization to higher loops} Our results in this section holds for one loop diagrams. For higher loop order there are simple examples where the wavefunction can also be written in the form of \eqref{amprepn} (see \ref{app:b} for a two loop example). We will leave a careful proof for \eqref{amprepn} for diagrams with any topology at arbitrary loop order to the future.

\section{From analyticity of amplitudes to analyticity of the wavefunction}\label{sec:analyticity}
In \cite{Salcedo2022}, when we studied the analyticity of the wavefunction, we arrived at the "energy conservation condition", i.e. singularities in the wavefunction corresponds to the vanishing of energy entering into a subgraph of a Feynman diagram. This approach has given us a set of singularities for the off-shell wavefunction: namely, the singularities must be located on the negative real axis of the complex plane of one of the off-shell external energies, say $\omega_1$. 

Our results from the previous section will allow us to say more about the analyticity of the wavefunction. We have shown that the one loop wavefunction can be divided into a total energy pole, whose residue is the amplitude, and as a remainder part. Therefore, we expect the following:
\begin{itemize}
    \item A subset of the singularities of the one loop wavefunction can be mapped to singularities of an amplitude. In fact, we will see that these singularities have the same interpretation as the corresponding singularities in amplitudes: namely, they correspond to cutting the same internal lines of a Feynman diagram and sending them "on-shell", i.e. demanding the corresponding propagator to diverge. We shall call these singularities "amplitude-type singularity"
    \item The remaining singularities have no analogue in amplitudes, and we shall call them "wavefunction-type singularities". Interestingly, in the one loop wavefunction, they correspond to cutting a single internal line (which gives no physical singularities for amplitudes). We shall have more to say about these singularities in the next section.
\end{itemize}

Computing the one loop wavefunction for a given diagram is significantly more difficult than the computation of amplitude for the same diagram (for reference, see Appendix A of \cite{Salcedo2022}). There is little hope to see the analyticity of a wavefunction simply by staring at the full wavefunction, except for the simplest cases. Therefore, we will make use of Landau analysis \cite{Landau:1959fi} to study the analytic structure of the wavefunction. 
\subsection{Landau analysis}
Let us write down the wavefunction in the following form:
\begin{equation}
    \psi_n=\int\left[\prod_i\frac{ds_i}{2\pi i}\right]\frac{R(\{\omega\},\{s\})}{D_K(\{\omega\},\{s\})}\int\prod_l\frac{d^Dp_l}{(2\pi)^D}\frac{1}{D_A(\{\bfp\},\{s\})}.\label{integral}
\end{equation}
This is merely a rewriting of \eqref{Wfnmaster}. To make it more clear, we have:
\begin{equation}
    \frac{R(\{\omega\},\{s\})}{D_K(\{\omega\},\{s\})}=\prod_a\tilde{D}_a.
\end{equation}
The main difference here is that on the left hand side we have gathered all the factors in the same denominator. More explicitly, 
\begin{align}
    D_K=\prod_a\prod_{\sigma_j=\pm}\left(\omega_a^2-\left(s_1+\sum_{j=2}^m \sigma_j s_j\right)^2-i\epsilon\right), \label{Dk}
\end{align}
and $R$ is a polynomial in $s$ and $\omega$. $D_A$ is simply the following product:
\begin{equation}
    D_A=\prod_i (s_i^2-\Omega_{pi}^2+i\epsilon).
\end{equation}

As an example, for a bubble diagram, we can start from \eqref{da2} and obtain:
\begin{equation}
    \tilde{D}_1\tilde{D}_2=\frac{16s_1^2s_2^2\omega_1\omega_2}{(\omega_1^2-(s_1+s_2)^2-i\epsilon)(\omega_1^2-(s_1-s_2)^2-i\epsilon)(\omega_2^2-(s_1+s_2)^2-i\epsilon)(\omega_2^2-(s_1-s_2)^2-i\epsilon)}.
\end{equation}
Here $D_K$ has the form \eqref{Dk}, and $R=16s_1^2s_2^2\omega_1\omega_2$. 

We can introduce a Feynman parameter $\lambda$ and write:
\begin{equation}
    \psi_n=\int_{0}^{\infty}d\lambda\int\left[\prod_i\frac{ds_i}{2\pi i}\right]\int\prod_l\frac{d^Dp_l}{(2\pi)^D}\frac{R(\{\omega\},\{s\})}{D(\{\bfp\},\{\omega\},\{s\})},
\end{equation}
\begin{equation}
    D(\{\bfp\},\{\omega\},\{s\})=D_K(\{\omega\},\{s\})+\lambda D_A(\{\bfp\},\{s\}).
\end{equation}

Let us write down the Landau equations. We have:
\begin{align}
    D=0&\rightarrow D_K+\lambda D_A=0, \\
    \frac{\partial D}{\partial \lambda}=0&\rightarrow D_A=0.
\end{align}
Let's interpret these two equations before writing down the rest. Since $D_A$ is just a product of internal propagators, demanding $D_A=0$ means sending a subset of propagators on shell (graphically this is simply cutting internal lines). At least one of the internal lines must be cut if we want a physical singularity. In addition, combining these two equations gives us $D_K=0$. From \eqref{Dk} we can see that this fixes the internal energies $s$ of the propagators in terms of external energies $\omega$. 

Let us write down the remaining equations. They are:
\begin{align}
    \frac{\partial D}{\partial s_i}=0&\rightarrow \frac{\partial D_K}{\partial s_i}+\lambda \frac{\partial D_A}{\partial s_i}=0\,\,\,\text{(for all $S_i$)},\\
    \frac{\partial D}{\partial \bfp_l}=0&\rightarrow\lambda\frac{\partial D_A}{\partial \bfp_l}=0.
\end{align}
The second equation, together with $D_A=0$, give us a modified version of the Landau equation for amplitudes. To see this, start with \eqref{integral}, then introduce Feynman parameters $\alpha_{pi}$ for $D_A$ before introducing the extra Feynman parameter $\lambda$. Then we get the following Landau equations involving $D_A$:
\begin{align}
    &|\bfp_i|^2=s_i^2-m_i^2,\\
    &\sum_{i\in l}\alpha_{pi}\bfp_i=0.
\end{align}
These are simply Landau equations for the amplitude with the same Feynman diagram, except we have modified $m_i^2$ to $s_i^2-m_i^2$. Of course, we still need to solve $\frac{\partial D}{\partial s_i}$ to fix $s_i$, but often we will find that knowing the solutions for the Landau equations in amplitudes will give us shortcuts to solving the Landau equations for the wavefunction.

In fact, we could even carry out the momentum integral first before writing down the Landau equation. This gives:
\begin{equation}
    D_A=\frac{\partial D_A}{\partial \alpha_{pi}}=0.
\end{equation}
 
\paragraph{Counting number of constraints} Before moving on to concrete examples, let us first ask an important question: are the Landau equations sufficient to fix the singularities? 

In general, we have $I+L+1$ integration variables which require fixing, ($I$ is the total number of internal lines, and $L$ is the total number of loops). We have $I+L+2$ Landau equations, so there are enough equations to fix all the integration variables, and we can always write down the singularities in terms of external kinematics. However, if we want to write down the singularity for one of the external energies $\omega_1$, quite often we will need to write it in terms of other external off-shell energies (say $\omega_2$) as well. This creates an obstacle when we study the singularities: for instance, suppose a singularity surface $\omega_1=-\omega_2$, naively we would assume that $\omega_2$ can take on any real negative value. However, once this energy is on-shell (say $\omega_2=\sqrt{k^2+4m^2}$), we realise that $\omega_1<-2m$. This is an example where information is lost when the other external energies are off-shell. Another example would be the anomalous threshold of the three site one loop diagram, where the singularity condition is expressed purely in terms of the mass of internal and external particles (see \eqref{anomalousth}). Without putting external energies on-shell, it would be hard to recognise the anomalous threshold since off-shell external energies have no explicit dependence on mass.

It is often helpful to supply additional information on these external energies. For example, we can put some of the external energies on-shell. We will see an example of this when we tackle the three site one loop diagram. 

\subsection{Example: two site loop}
Let's start with the two site loop (figure \ref{fig:2siteclean}). For simplicity we will always consider the internal lines to have the same mass $m$.

Before we carry out Landau analysis, let us set our expectations for the result. We know that the recursion expression for the wavefunction is given by \cite{Arkani-Hamed:2017fdk}:
\begin{equation}
    \psi_2=\frac{1}{\omega_T}\int_{\bfp}\frac{1}{(\omega_1+\Omega_{p1}+\Omega_{p2})(\omega_2+\Omega_{p1}+\Omega_{p2})}\left[\frac{1}{\omega_1+\omega_2+2\Omega_{p1}}+\frac{1}{\omega_1+\omega_2+2\Omega_{p2}}\right].
\end{equation}
Previously in \cite{Salcedo2022} we found the following singularities:
\begin{itemize}
    \item $\omega_1=-\sqrt{k^2+4m^2}$. We will see that this corresponds to an amplitude-type singularity.
    \item $\omega_1=-\omega_2-2m$. We will see that this corresponds to a wavefunction-type singularity.
\end{itemize}

Let us write down the expression for the wavefunction coefficient of the two site loop again:
\begin{equation}
	\psi_2=\int\frac{ds_1}{2\pi i}\int\frac{ds_2}{2\pi i}\frac{16 s_1^2 s_2^2\omega_1\omega_2}{S_{1+}S_{1-}S_{2+}S_{2-}}\int_{\bfp}\frac{1}{(s_1^2-\Omega_{p_1}^2+i\epsilon)(s_2^2-\Omega_{p_2}^2+i\epsilon)}
\end{equation}
\begin{align}
	S_{1\pm}&=\omega_1^2-(s_1\pm s_2)^2-i\epsilon,\\
	S_{2\pm}&=\omega_2^2-(s_1\pm s_2)^2-i\epsilon.
\end{align}

The convention for the internal momenta is $\bfp_1+\bfp_2=-\bfk$.

To simplify our results:
\begin{itemize} 
    \item We will not look at cases where we have $S_{1+}=0$ and $S_{1-}=0$ simultaneously. This is because they both come from the sum in $\tilde{D}_1$ (see \eqref{tilD1}). We only need to send an individual term in the sum to infinity, which corresponds to sending one of $S_{1+}$ or $S_{1-}$ to infinity (and also picking a sign, for example picking $\omega_1=-s_1+s_2$ instead of $\omega_1=s_1+s_2$). 
    \item We will also impose $\omega_2>0$, this will restrict the singularities for $\omega_1$ to be on the negative real axis.
\end{itemize}
We will look at the case where $S_{1+}=0$, and potentially $S_{2-}=0$. The case where we have $S_{1-}=0$ (and $S_{2+}=0$) are easily generalisations. 

Let us introduce Feynman parameters for both $D_K$ and $D_A$. The integral now has the following form:
\begin{multline}
	\psi_2=\int\frac{ds_1}{2\pi i}\int \frac{ds_2}{2\pi i}\frac{16 s_1^2 s_2^2 \omega_1\omega_2}{S_{1_-}S_{2_+}}\int_{0}^{1}d\alpha_{1_+} \int_{0}^{1}d\alpha_{2_-}\int_{0}^{1}d\alpha_{p_1}\int_{0}^{1}d\alpha_{p_2}\\\times\int_{\bfp}\frac{\delta(1-\alpha_{1_+}-\alpha_{2_-})\delta(1-\alpha_{p_1}-\alpha_{p_2})}{(\alpha_{1_+}(\omega_1^2-(s_1+s_2)^2)+\alpha_{2_-}(\omega_2^2-(s_1-s_2)^2))^2(\alpha_{p_1}(s_1^2-\Omega_{p_1}^2)+\alpha_{p_2}(s_2^2-\Omega_{p_2}^2))^2}.
\end{multline}
We did not introduce Feynman parameters for $S_{1-}$ and $S_{2+}$ since we will never send them to zero. Just like Landau equations for amplitudes, we can do the momentum integral first. This will allow us to write down Landau equations in terms of Feynman parameters $\alpha$, $\lambda$, and the internal energies $s$. This gives an equation of the form \eqref{integral}, with

\begin{equation}
	D_A=\frac{\alpha_{p_1}\alpha_{p_2}}{\alpha_{p_1}+\alpha_{p_2}}k^2-\alpha_{p_1}(s_1^2-m^2)-\alpha_{p_2}(s_2^2-m^2).
\end{equation}
\begin{equation}
	D_K=-\alpha_{1_+}(\omega_1^2-(s_1+s_2)^2)-\alpha_{2-}(\omega_2^2-(s_1-s_2)^2).
\end{equation}

\begin{figure}[h!]
    \centering
    \includegraphics[scale=1.2]{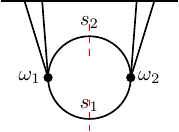}
    \caption{The amplitude-type singularity. The lines cut by the red dashed lines are the propagators which are on-shell.}
    \label{fig:2site2cut}
\end{figure}

\paragraph{Amplitude-type singularity}
As an example, let us solve the Landau equations with $S_{1_+},S_{p1},S_{p2}= 0$. Since both $S_{p1}$ and $S_{p2}$ are zero, we are cutting both propagators (see figure \ref{fig:2site2cut}). Let's see if this singularity can be mapped to a singularity in amplitudes.

The Landau equation reads:
\begin{align}
	\frac{\partial D}{\partial\alpha_{p_1}}=0&\rightarrow\frac{\alpha_{p_2}^2}{(\alpha_{p_1}+\alpha_{p_2})^2}k^2=s_1^2-m^2,\label{dDdap1}\\
	\frac{\partial D}{\partial\alpha_{p_2}}=0&\rightarrow\frac{\alpha_{p_1}^2}{(\alpha_{p_1}+\alpha_{p_2})^2}k^2=s_2^2-m^2,\\
	\frac{\partial D}{\partial\alpha_{1_+}}=0&\rightarrow\omega_1^2=(s_1+s_2)^2,\\
	\frac{\partial D}{\partial s_i}=0&\rightarrow \lambda\alpha_{p_1}s_1=\lambda\alpha_{p_2}s_2=(s_1+s_2).
\end{align}
Also, naturally we demand $\alpha_{p_1}+\alpha_{p_2}=1$. Since $S_{2-}\neq 0$, $\alpha_{2_-}$ must be zero and this naturally means $\alpha_{1_+}=1$.

The first two equations give:
\begin{equation}
	\alpha_{p_2}^2(s_2^2-m^2)=\alpha_{p_1}^2(s_1^2-m^2).
\end{equation}
Therefore, using the last Landau equation, we obtain:
\begin{equation}
	\frac{s_1^2-m^2}{s_2^2-m^2}=\frac{s_1^2}{s_2^2}.
\end{equation}
The solution for this is $s_1=\pm s_2$ unless $s_1=\pm m$ or $s_2=\pm m$. However, if we set $s_1=\pm m$, the \eqref{dDdap1} equation implies that $k=0$ or $\alpha_{p2}=0$. But if $\alpha_{p2}=0$ that would imply $s_1=-s_2$. This would mean that if $s_1$ is in the lower half plane, $s_2$ is in the upper half plane, so either $s_1$ or $s_2$ are not in the physical integration region. Therefore we conclude that if $s_1=\pm m$ we must have $k=0$. 

First consider $k\neq 0$. Both $s_1$ and $s_2$ must lie in the lower half complex plane. One can easily show that we have $s_1=s_2$, so this gives us the singularity $\omega_1^2=4s_1^2$. It is also straightforward to see $\lambda=4$. Using this, we get:
\begin{equation}
	\frac{k^2}{4}=s_1^2-m^2
\end{equation}
This gives us the $\omega_1=-\sqrt{k^2+4m^2}$ singularity. 

For $k=0$ we can easily see that this implies $\omega_1=-2m$. But this is just a special case of $\omega_1=-\sqrt{k^2+4m^2}$, and so we landed on the same singularity.

How is this singularity connected to an amplitude singularity? We know that for the same diagram, an amplitude has the singularity $s=4m^2$ (here $s$ is the Mandelstahm variable). But $s=\omega_1^2-k^2$ (since $\omega_1$ is the total energy entering the vertex and $\bfk$ is the total momentum entering the vertex), hence we have $\omega_1^2=k^2+4m^2$. Therefore this $\omega_1=-\sqrt{k^2+4m^2}$ singularity is simply the same singularity as $s=4m^2$, written in terms of the off-shell energy variables.

\paragraph{Amplitude-type singularity for massless particles}
It is also helpful to look at singularities of massless internal particles. Let us solve the same Landau equations, but with $m=0$.
The Landau equation reads:
\begin{align}
	\frac{\partial D}{\partial\alpha_{p_1}}=0&\rightarrow\frac{\alpha_{p_2}^2}{(\alpha_{p_1}+\alpha_{p_2})^2}k^2=s_1^2-m^2,\\
	\frac{\partial D}{\partial\alpha_{p_2}}=0&\rightarrow\frac{\alpha_{p_1}^2}{(\alpha_{p_1}+\alpha_{p_2})^2}k^2=s_2^2-m^2,\\
	\frac{\partial D}{\partial\alpha_{1_+}}=0&\rightarrow\omega_1^2=(s_1+s_2)^2,\\\label{dDda1}
	\frac{\partial D}{\partial s_i}=0&\rightarrow \lambda\alpha_{p_1}s_1=\lambda\alpha_{p_2}s_2=(s_1+s_2).
\end{align}
The solution for this is quite straightforward: the first two equations imply $\alpha_{p2}k=s_1$ and $\alpha_{p1}k=s_2$. From this we can easily conclude that $s_1+s_2=(\alpha_{p1}+\alpha_{p2})k=k$. Substituting this into \eqref{dDda1} gives $\omega_1=-k$, which is the massless version of the amplitude-type singularity we just found.

We can easily write down the solutions to the Feynman parameters and the internal energies. Just take:
\begin{align}
    &s_1=s_2=\frac{k}{2},\\
    &\alpha_{p1}=\alpha_{p2}=\frac{1}{2},\\
    &\lambda=4.
\end{align}
It is easy to show by substitution that this solves the Landau equations.

\begin{figure}[h!]
    \centering
    \includegraphics[scale=1.2]{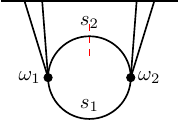}
    \caption{Wavefunction-type singularity. Only one propagator is on-shell}
    \label{fig:2site1cut}
\end{figure}

\paragraph{Wavefunction-type singularity}
Now let us consider the case where $S_{1+},S_{2-},S_{p2}=0$, but we have $S_{p1}\neq 0$. Notice that only one internal line is cut here (see figure \ref{fig:2site1cut}). For amplitude there is no singularity for cutting one internal line. However, for the wavefunction there is a singularity. In fact, for massless particles, it is a physical singularity for any external kinematics.

Let us write down the Landau equation. They are
\begin{align}
	\frac{\partial D}{\partial\alpha_{1_+}}=0&\rightarrow\omega_1^2=(s_1+s_2)^2,\\
	\frac{\partial D}{\partial\alpha_{2_-}}=0&\rightarrow\omega_2^2=(s_1-s_2)^2,\\
	\frac{\partial D}{\partial \alpha_{p_2}}=0&\rightarrow 0=s_2^2-m^2,\\
	\frac{\partial D}{\partial s_1}=0&\rightarrow 0=\alpha_{1_+}(s_1+s_2)+\alpha_{2_-}(s_1-s_2),\\
	\frac{\partial D}{\partial s_2}=0&\rightarrow \lambda s_2=\alpha_{1_+}(s_1+s_2)-\alpha_{2_-}(s_1-s_2)
\end{align}
The third equation gives us $s_2=m$. We throw away $s_2=-m$: if we restore the $i\epsilon$ in the solution, this is $s_1=-m+i\epsilon$, and is not included in the integration region, i.e. the lower half complex plane.

In addition, we have:
\begin{align}
	\omega_1+s_1+s_2&=0,\\
	\omega_2-s_1+s_2&=0.
\end{align}
Use this to eliminate $s_1$, we obtain $\omega_1+\omega_2=-2m$. This singularity is a total energy singularity, and quite obviously have no analogue in amplitudes. 

Is this singularity physical? Let us compute the Feynman parameters. We can show that:
\begin{equation}
    \alpha_{2-}=-\frac{\omega_1}{2m}.
\end{equation}
But $\omega_2=-\omega_1-2m\geq 0$, so we get $\alpha_{2-}\geq 1$. Therefore this singularity may not be visible for any general kinematics! Indeed this is consistent with the observation that this singularity is in fact invisible outside the soft limit (see the appendix of \cite{Salcedo2022}). However it is not entirely clear how this emerges from the Landau analysis picture, and we will leave a detailed study of this for the future.

\paragraph{Wavefunction-type singularity for massless particles}
Things are more clear in the case of massless particles. The Landau equations read:
\begin{align}
	\frac{\partial D}{\partial\alpha_{1_+}}=0&\rightarrow\omega_1^2=(s_1+s_2)^2\rightarrow\omega_1+s_1+s_2=0,\\
	\frac{\partial D}{\partial\alpha_{2_-}}=0&\rightarrow\omega_2^2=(s_1-s_2)^2\rightarrow \omega_2-s_1+s_2=0,\\
	\frac{\partial D}{\partial \alpha_{p_2}}=0&\rightarrow 0=s_2^2,\\
	\frac{\partial D}{\partial s_1}=0&\rightarrow 0=\alpha_{1_+}(s_1+s_2)+\alpha_{2_-}(s_1-s_2),\\
	\frac{\partial D}{\partial s_2}=0&\rightarrow \lambda s_2=\alpha_{1_+}(s_1+s_2)-\alpha_{2_-}(s_1-s_2)
\end{align}
Immediately we have $s_2=0$, and we have $\omega_1+\omega_2=0$. This is indeed the massless limit of the wavefunction-type singularity found above.

Unlike the massive case, this singularity is present for any external kinematics. Indeed, if we take $s_1=0$, we can easily see that the Landau equations can always be satisfied, with $\alpha_{1+},\alpha_{2-}>0$. 
\paragraph{Where does the wavefunction-type singularity come from?}
Before we move on it is helpful to see how the wavefunction-type singularity emerges from the expression we have obtained, i.e. from \eqref{psi2}. Now clearly the singularity cannot come from the amplitude part of the wavefunction, therefore it must be a singularity from the remainder terms.

Upon inspection, we find that the singularity actually emerges from $\psi_2^{\text{FB}}$, i.e. the mixing term between Feynman propagator and the boundary term. We have shown that this term can be written as \eqref{FBpsi2}. If we try to study the singularity associated with $\omega_1+\omega_2+2\Omega_{p1}=0$, we find the wavefunction-type singularity from above.

The terms $\psi_2^{\text{FF}}$ and $\psi_2^{\text{BB}}$ do not provide any contribution to the wavefunction-type singularity. Only the $\omega_1=-\sqrt{k^2+4m^2}$ singularity is present for those terms if we study the analyticity of these terms individually.

\subsection{Example: three site loop}
Let us now look at the one loop three site graph. For amplitudes, this graph has two types of singularities. The first type is the normal threshold, where two internal lines are cut. This type of cut is associated with unitarity in the optical theorem. The second type is the anomalous threshold, where all three internal lines are cut. 

Since we have shown that the wavefunction can be written in terms of amplitude plus remainder terms, we would expect singularities for the amplitude to be present in the wavefunction as well, including the anomalous threshold. We will see that it is indeed the case. In addition, we will find that there are wavefunction-type singularities again, where only a single internal line is cut.

For the one loop three site graph, the wavefunction coefficient can be written as:
\begin{equation}
	\psi_3=\int\frac{ds_1}{2\pi i}\int\frac{ds_2}{2\pi i}\int\frac{ds_3}{2\pi i}\frac{512\omega_1\omega_2\omega_3s_1^2s_2^2s_3^2}{S_{1+}S_{1-}S_{2+}S_{2-}S_{3+}S_{3-}}\int_{\bfp}\frac{1}{(s_1^2-\Omega_{p_1}^2)(s_2^2-\Omega_{p_2}^2)(s_3^2-\Omega_{p_3}^2)}.
\end{equation}
\begin{align}
	&S_{1\pm}=\omega_1^2-(s_1\pm s_3)^2,\\
	&S_{2\pm}=\omega_2^2-(s_1 \pm s_2)^2,\\
	&S_{3\pm}=\omega_3^2-(s_2\pm s_3)^2.
\end{align}
Also we have $\bfp_1=\bfp$, $\bfp_2=(\bfp+\bfk_2)$, $\bfp_3=(\bfp-\bfk_1)=(\bfp+\bfk_2+\bfk_3)$.

Let's write everything in terms of Feynman parameters and perform the $\bfp$ integral first. We obtain:
\begin{equation}
	\psi_3=\int_{0}^{\infty}d\lambda\int \prod_{i=1}^{3}\frac{ds_i}{2\pi i}\int_0^1 \prod_{j=1\pm,2\pm,3\pm} d\alpha_j \int_{0}^{1}\prod_{l=p_1,p_2,p_3}d\alpha_{l}\frac{F(\{\alpha\})\delta(1-\sum\alpha)}{D(\{\alpha\},s_1,s_2,s_3)^{9-d/2}},
\end{equation}
where we have:
\begin{equation}
	D_k=\frac{\alpha_{p_1}\alpha_{p_2}k_2^2+\alpha_{p_2}\alpha_{p_3}k_3^2+\alpha_{p_1}\alpha_{p_3}k_1^2}{\sum\alpha_{p_i}}-\sum_{i=1}^{3}\alpha_{p_i}(s_i^2-m^2),
\end{equation}
as well as
\begin{equation}
	D_A=-\sum_{i=1}^{3}\alpha_{i\pm}S_{i\pm}.
\end{equation}
$F({\alpha})$ is just polynomial in $\alpha$, and its exact form will not be important for us.
\paragraph{Amplitude-type singularity: normal threshold}
Let us first study solutions corresponding to "normal thresholds". These are solutions where $S_{p_i}\neq0$ for one $p_i$, and they can be used to recover certain energy conservation poles.

\begin{figure}[h!]
    \centering
    \includegraphics[scale=1.2]{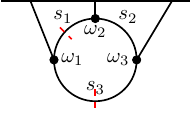}
    \caption{Amplitude-type singularity, with the internal line $s_1$ and $s_3$ being cut.}
    \label{fig:3sites2}
\end{figure}

As an example let us consider setting $S_{p_2}\neq0$, which requites $\alpha_{p_2}=0$. In amplitude terms we are looking at a singularity which comes from cutting $s_1$ and $s_3$ (see figure \ref{fig:3sites2}). Then we have the following equations from $D_1$:
\begin{align}
	\frac{\alpha_{p_3}^2k_1^2}{(\alpha_{p_1}+\alpha_{p_3})^2}&=s_1^2-m^2,\\
	\frac{\alpha_{p_1}^2k_1^2}{(\alpha_{p_1}+\alpha_{p_3})^2}&=s_3^2-m^2
\end{align}
At this point if we choose $S_{1+}=0$ we can pretty much repeat the analysis we did for the two site graph and obtain $\omega_1=-\sqrt{k^2_1+4m^2}$.

\begin{figure}[h!]
    \centering
    \begin{subfigure}[b]{0.47\textwidth}
        \centering
        \includegraphics[scale=1.2]{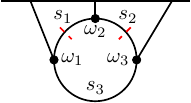}
    \end{subfigure}
    \begin{subfigure}[b]{0.47\textwidth}
        \centering
        \includegraphics[scale=1.2]{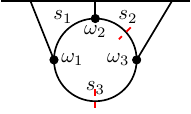}
    \end{subfigure}
    \caption{Two other singularities for the wavefunction which are also amplitude-type singularities. They are $\omega_1+\omega_3=\sqrt{k_2^2+4m^2}$ (for the left figure) and $\omega_1+\omega_2=-\sqrt{k_3^2+4m^2}$ (for the right figure)}
    \label{fig:3site2cuts2}
\end{figure}

\begin{figure}[h!]
    \centering
    \includegraphics[scale=1.5]{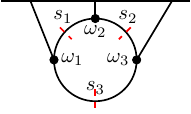}
    \caption{The anomalous amplitude-type singularity. To access this singularity we also need specific external kinematics, which are given by \eqref{anomalousbegin}-\eqref{anomalousend}.}
    \label{fig:Anomalous}
\end{figure}

However, we can make a more interesting choice. Suppose we choose $S_{2_-}=S_{3_+}=0$ instead. (I will relate this to a singularity in $\omega_1$ by permutation). Then we get the following equations from $D_2$:
\begin{align}
	\lambda\alpha_{p_1}s_1&=\alpha_{2_-}(s_1-s_2),\\
	\lambda\alpha_{p_3}s_3&=\alpha_{3_+}(s_3+s_2),\\
	0&=-\alpha_{2_-}(s_1-s_2)+\alpha_{3_+}(s_3+s_2).
\end{align}
This can be rearranged into the following:
\begin{equation}
	\frac{\alpha_{p_1}s_1}{\alpha_{p_3}s_3}=\frac{\alpha_{2_-}(s_1-s_2)}{\alpha_{3_+}(s_3+s_2)}=1.
\end{equation}
Combine this with the equations from $D_1$, we get:
\begin{equation}
	\frac{s_1^2}{s_3^2}=\frac{s_1^2-m^2}{s_3^2-m^2}.
\end{equation}
Then we can just following the derivation as in the two site graph to get:
\begin{align}
	\omega_2&=-\sqrt{\frac{k_1^2}{4}+m^2}+s_2,\\
	\omega_3&=-\sqrt{\frac{k_1^2}{4}+m^2}-s_2.
\end{align}
This combines to give $\omega_2+\omega_3=-\sqrt{k_1^2+4m^2}$ singularity. This is related by permutation to $\omega_1+\omega_2=-\sqrt{k_3^2+4m^2}$ and $\omega_1+\omega_3=\sqrt{k_2^2+4m^2}$ singularity (see figure \ref{fig:3site2cuts2}), both of which are energy conserving poles.
\paragraph{Amplitude-type singularity: Anomalous threshold}
For amplitudes there is an anomalous threshold that corresponds to sending all three internal lines on-shell (see figure \ref{fig:Anomalous}). The existence of this threshold depends on the external kinematics: namely, two of the external four momenta have to be equal. It is known that the threshold is only physical when the masses of the external particles satisfy $M^2>2m^2$ where $m^2$ is the mass of the internal particles \cite{Anomalous,Anomalous2,Anomalous3,Anomalous4, Correia:2020xtr, Correia:2022dcu, Sashanotes}. This is lower than the expected two particle threshold $4m^2$, hence the name anomalous threshold. This threshold is given by:
\begin{equation}
    s=4m^2-\frac{(M^2-2m^2)^2}{m^2}, \label{anomalousth}
\end{equation}
where $s$ is the four momentum for one of the external lines (in the diagram it would be the line attached to $\omega_1$) vertex.

We would like to show that such a threshold exists for the wavefunctions as well. We expect that the singularity only appears for specifically chosen kinematics just like amplituides. To make life easier we will choose the the following kinematics:
\begin{align}
	\bfk_2&=-\bfk_3,\label{anomalousbegin}\\
	|\bfk|&=k,\\
	\bfk_1&=0,\\
	\omega_2&=\omega_3=\sqrt{k^2+M^2},\\
	\omega_1^2&=4\omega_2^2\label{anomalousend}.
\end{align}
The last line enforces the total energy to be zero, and we know in this limit the wavefunction reduces to the amplitude, hence we should expect the anomalous threshold to show up.

Now let's specify which $S_{i\pm}$ we send to zero. We make the following choice:
\begin{align}
	S_{1+}&=\omega_1^2-(s_1+s_3)^2=0,\\
	S_{2+}&=\omega_2^2-(s_1+s_2)^2=0,\\
	S_{3+}&=\omega_3^2-(s_3+s_2)^2=0.
\end{align}
So now we have:
\begin{align}
	\omega_1&=s_1+s_3,\\
	\omega_2&=s_1+s_2,\\
	\omega_3&=s_3+s_2.
\end{align}
as well as:
\begin{align}
	\alpha_{1_+}(s_1+s_3)+\alpha_{2_+}(s_1+s_2)&=\lambda\alpha_{p1}s_1,\label{1landauD2}\\	\alpha_{2_+}(s_1+s_2)+\alpha_{3_+}(s_3+s_2)&=\lambda\alpha_{p_2}s_2,\label{2landauD2}\\
	\alpha_{3_+}(s_3+s_2)+\alpha_{1_+}(s_1+s_3)&=\lambda\alpha_{p_3}s_3\label{3landauD2},
\end{align}
Also from $D_A$ we have:
\begin{align}
	\frac{(\alpha_{p_1}+\alpha_{p_3})^2k_2^2-\alpha_{p_1}\alpha_{p_3}k_1^2}{(\alpha_{p_1}+\alpha_{p_2}+\alpha_{p_3})^2}-(s_2^2-m^2)&=0,\label{1LandauD1}\\
	\frac{\alpha_{p_2}^2k_2^2+\alpha_{p_3}(\alpha_{p_2}+\alpha_{p_3})k_1^2}{(\alpha_{p_1}+\alpha_{p_2}+\alpha_{p_3})^2}-(s_1^2-m^2)&=0\label{2LandauD1},\\
	\frac{\alpha_{p_2}^2k_2^2+\alpha_{p_1}(\alpha_{p_2}+\alpha_{p_1})k_1^2}{(\alpha_{p_1}+\alpha_{p_2}+\alpha_{p_3})^2}-(s_3^2-m^2)&=0\label{3LandauD1}.
\end{align}

Because $\omega_2=\omega_3$ this immediately gives $s_1=s_3$. Now we can use \eqref{2LandauD1} and \eqref{3LandauD1} to obtain:
\begin{equation}
	\alpha_{p_1}(\alpha_{p_2}+\alpha_{p_1})=\alpha_{p_3}(\alpha_{p_2}+\alpha_{p_3}),
\end{equation}
which implies $\alpha_{p_1}=\alpha_{p_3}$. We also use \eqref{1landauD2} and \eqref{3landauD2} (and $s_1=s_3$) to show:
\begin{equation}
	\alpha_{2_+}=\alpha_{3_+}.
\end{equation}

Now using $\omega_1^2=4\omega_2^2$, we obtain:
\begin{equation}
	4(s_1+s_2)^2=4s_1^2,
\end{equation}
so $s_2=-2s_1$. This also gives $\omega_1=-2\omega_2$.

By using \eqref{1LandauD1} and \eqref{2LandauD1} (\eqref{3LandauD1} is the same as \eqref{2LandauD1}) and eliminating the Feynman parameters, we get:
\begin{equation}
	(s_2^2-s_1^2+k^2)^2=4k^2(s_2^2-m^2).
\end{equation}
Usually this is as far as we can go, but because now we fixed $\omega_1^2=4\omega_2^2$ we found $s_2=-2s_1$, so we can further simplify this. Eventually we get:
\begin{equation}
	\frac{M^4}{m^2}=(\omega_1^2-4M^2),
\end{equation}
or
\begin{equation}
	\omega_1^2=4m^2-\frac{(M^2-2m^2)^2}{m^2},
\end{equation}
which is exactly the value for anomalous threshold in the amplitude case. Hence we have demonstrated the existence of anomalous threshold in the wavefunction as well.

We can also compute the Feynman parameters explicitly to show that this is indeed a physical singularity. For instance,
\begin{equation}
    \alpha_{p_2}=\frac{M^2-2m^2}{M^2}.
\end{equation}
For the singularity to be real, $0<\alpha_{p_2}<1$. Hence We must have $M^2>2m^2$, which is indeed a criteriom for anomalous threshold.

We could also compute the other Feynman parameters, which are found to be:
\begin{align}
    &\alpha_{2_+}=\frac{4M^2-4m^2}{10M^2-18m^2},\\
    &\lambda=\frac{4M^2}{10M^2-18m^2}.
\end{align}
For $M^2>2m^2$ we have $0<\alpha_{2+}<1$, as well as $\lambda>0$, hence this is not a spurious singularity.

Which singularity does this correspond to in the recursion expression? It turns out that it corresponds to this condition:
\begin{equation}
	\omega_1+\Omega_{p_1}+\Omega_{p_3}=0,
\end{equation}
along with:
\begin{equation}
	\omega_1+\omega_2+\omega_3=0.
\end{equation}
This is reflecting an interesting fact about studying singularities using the recursion expression. Naively if we start with the expression $\omega_1+\Omega_{p_1}+\Omega_{p_3}=0$ and try to minimize it, it is tempting to simply write down the most obvious solution (which is $\omega_1=-\sqrt{k^2+4m^2}$), and we would be led to incorrect conclusions about the analytic structure (for instance, that the singularity must first appear at $\omega_1<-2m$). What our result shows is that at particular kinematics, there may be other minimum solutions that corresponds to additional singularities. For instance in this case we found a singularity that can live in the range $-2m<\omega_1<\sqrt{2}m$. 

This is one of the main advantages of studying the analytic structure of the wavefunction using the amplitude representation rather than using the recursion expression: we have a much better understanding of the analytic structure of amplitudes, and this formalism give us a mapping between singularities of amplitudes and the wavefunction. If we also have a good understanding of how wavefunction-type singularities emerge (for example, in one loop cases they are singularities corresponding to cutting only one internal line) then we have a complete catalogue of the singularities present in the wavefunction.

\begin{figure}[h!]
    \centering
    \includegraphics[scale=1.2]{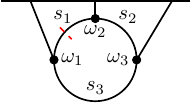}
    \caption{Wavefunction-type singularity for the three site one loop graph.}
    \label{fig:3site1cut}
\end{figure}

\paragraph{Wavefunction-type singularity}
One can also show that the wavefunction-type singularities are also present. To see this, let us just cut one internal line (see figure \ref{fig:3site1cut}), and pick:
\begin{align}
    &S_{1+}=\omega_1^2-(s_1+s_3)^2=0,\\
    &S_{2-}=\omega_2^2-(s_1-s_2)^2=0,\\
    &S_{3-}=\omega_3^2-(s_2-s_3)^2=0,\\
    &S_{p1}=s_1^2-\Omega_{p1}^2=0.
\end{align}
Then one can show, similar to the two site graph case, that we have $\omega_1+\omega_2+\omega_3=-2m$ as a singularity. 

Once again it is instructive to see which term in $\psi_3$ give rise to this wavefunction-type singularity. Upon inspection, we can find that it is \eqref{psi3FFB} which has this singularity. Specifically, if we send $\omega_1+\Omega_{p_1}-s_3=0$, $\omega_2+\Omega_{p_1}+s_2=0$ as well as $\omega_3-s_2+s_3=0$, we will land on $\omega_1+\omega_2+\omega_3=-2m$. Interestingly, $\psi_3^{\text{FBB}}$ does not seem to give rise to a wavefunction-type singularity.

\section{From the wavefunction to in-in correlators via analyticity}\label{sec:correlators}
While it is important to study wavefunction coefficients, the ultimate goal is to understand the in-in correlators of fields. The in-in correlator is simply:
\begin{equation}
    \langle\phi(k_1)\phi(k_2)\dots\phi(k_n)\rangle=\,_\text{in}\langle 0|\phi(k_1)\phi(k_2)\dots\phi(k_n)|0\rangle_\text{in},
\end{equation}
i.e. field operators (usually at the same time) sandwiched between two in-vacuum state (i.e. vacuum located at the far past).  In general, to pass from the wavefunction to correlators, one uses the Born rule, which schematically reads:
\begin{equation}
    \langle\phi(k_1)\dots\phi(k_n)\rangle=\int D\phi\,\phi(k_1)\dots\phi(k_n)|\Psi|^2.
\end{equation}

In cosmology it is the in-in correlators that are linked with observables such as the CMB and large scale structure. Therefore it would also be helpful to understand whether the analytic structure of in-in correlators is the same as the analytic structure of wavefunctions.

It was noted that for massless particles, when we go from the wavefunction to correlators, certain singularities cancel \cite{Lee:2023jby}. As an example, consider the one site loop diagram, which looks like this:

\begin{figure}[h!]
    \centering
    \includegraphics[scale=1.2]{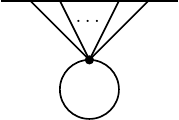}
    \caption{One site loop diagram}
    \label{fig:1site}
\end{figure}

The wavefunction has the form:
\begin{equation}
    \psi_1=\int_{\bfp}\frac{1}{\omega_T(\omega_T+2p)}=\frac{\omega_T}{16 \pi^2}\log(\omega_T).
\end{equation}
This wavefunction has a branch cut at $\omega_T=0$. However, one can show that the correlator is simply:
\begin{equation}
    B_1=\int_{\bfp}\frac{1}{2p}=0,
\end{equation}
if we use dimensional regularization. Obviously the correlator cannot have any branch cuts.

There are similar cancellations that occur for the two site loop. For massless particles we know the wavefunction for the graph to be the following (if we use cut-off regularization)\cite{Salcedo2022}:
\begin{multline}
	\psi_{2}(\omega_1,\omega_2,k)=\frac{1}{8\pi^2\omega_T}\left[\frac{\omega_2\log\left(\frac{\omega_1+k}{\Lambda}\right)-\omega_1\log\left(\frac{\omega_2+k}{\Lambda}\right)}{\omega_1-\omega_2}\right.\\
	-\left.\frac{\omega_{T} }{2k}\left(\frac{1}{2}\log^2\left(\frac{\omega_1+k}{\omega_2+k}\right)+\frac{\pi^2}{6}+\text{Li}_2\left(\frac{k-\omega_2}{k+\omega_1}\right)+\text{Li}_2\left(\frac{k-\omega_1}{k+\omega_2}\right)\right)\right] \; . 
\end{multline}
If we try to compute the in-in correlator, we get the following contributions: one from the time ordered propagators (where we use the $G_{++}$ and $G_{--}$ propagators):
\begin{equation}
	B_{2A}=\frac{1}{8\pi^2\omega_T}\left[\log(\frac{\omega_1+k}{\Lambda})+\log(\frac{\omega_2+k}{\Lambda})\right].
\end{equation}
There is also the out-of-time-ordered propagators (where we use the $G_{+-}$ and $G_{-+}$ propagators), which gives:
\begin{equation}
	B_{2B}=\frac{1}{\omega_1-\omega_2}\frac{-1}{8\pi^2}\left[-\log(\frac{\omega_1+k}{\Lambda})+\log(\frac{\omega_2+k}{\Lambda})\right].
\end{equation}
The in-in correlator is simply the sum of these two terms.

Notice that there is an important difference between the singularities of the wavefunction and of correlators:
\begin{itemize}
    \item For the wavefunction, both the amplitude-type singularity (given by $\omega_1=-k$) and the wavefunction-type singularity (given by $\omega_1+\omega_2=0$) are present. The wavefunction-type singularity is a total energy branch point.
    \item For the correlator, only the amplitude-type singularity (given by $\omega_1=-k$) is present. The wavefunction-type singularity is absent.
\end{itemize}

In both cases the wavefunction-type singularity are total energy branch points, and they are absent in the corresponding in-in correlator. This is particularly interesting since at one loop, wavefunction-type singularities all seem to be total energy branch points. In fact, one can show that the wavefunction-type singularity for an $n$ site one loop wavefunction is a total energy singularity. To see this, we use Landau analysis once again. The $n$ site one loop wavefunction is given by:
\begin{equation}
    \psi_n=\int\left[\frac{ds_1}{2\pi i}\dots \frac{ds_n}{2\pi i}\right]\frac{F(\{\omega\},\{s\})}{S_{1+}S_{1-}S_{2+}S_{2-}\dots S_{n+}S_{n-}}\int_{\bfp}\prod_{j=1}^{n}\frac{1}{s_j^2-\Omega_{pj}^2},
\end{equation}
where $S_{n\pm}=\omega_n^2-(s_n \pm s_{n-1})^2$. We can make use of Landau analysis again to find the wavefunction-type singularity. To do so, we only cut one line, since any other way of cutting should give us either normal or anomalous amplitude-type singularity. This is given by:
\begin{align}
    &S_{1+}=\omega_1^2-(s_1+s_{n})^2=0,\\
    &S_{2-}=\omega_2^2-(s_1-s_2)^2=0,\\
    &S_{3-}=\omega_3^2-(s_2-s_3)^2=0,\\
    &\,\,\,\,\,\,\,\,\,\,\,\,\,\,\,\,\,\dots\\
    &S_{n-}=\omega_n^2-(s_{n-1}-s_n)^2=0,\\
    &S_{p1}=s_1^2-\Omega_{p1}^2=0.
\end{align}
It is straightforward to show that this just gives $\omega_T=-2m$ as a singularity. 

At this point, we can apply the Cosmological KLN theorem \cite{Agui-Salcedo:2023wlq}, which is the following: \textit{total energy branch points in the wavefunction from loop integration are absent in the corresponding in-in correlator}. Since wavefunction-type singularities are all total energy branch points, we conclude that one loop in-in correlators do not possess any wavefunction-type singularities.

It is enticing to make the following conjecture, which extends our one loop result to all loop orders:

\begin{quote}
\centering
    \textit{Wavefunction-type singularities are always absent in in-in correlators.}
\end{quote}

This conjecture is in fact not surprising. Amplitudes are in-out correlators, i.e. field operators are sandwiched between vacuum in the asymptotic past (the in-vacuum) and asymptotic future (the out-vacuum). In flat space in-vacuum and out-vacuum are simply related by a phase \cite{kamenev_2011}. Assuming we have thermal equilibrium, we obtain:
\begin{equation}
    \,_\text{in}\langle 0|\phi(k_1)\phi(k_2)\dots\phi(k_n)|0\rangle_\text{in}=e^{i\theta}\,_\text{out}\langle 0|\phi(k_1)\phi(k_2)\dots\phi(k_n)|0\rangle_\text{in}.
\end{equation}
Since in-in correlators are related to amplitudes by a phase, one would not expect them to have different analytic structures. In other words, one should expect a one-to-one mapping between singularities of amplitudes to in-in correlators. However at this point our understanding of wavefunction-type singularities is still rather primitive, so we will leave a more careful study of this beyond one loop to the future.

\section{Conclusion and Outlook}\label{sec:outlook}
In this paper we have written down the off-shell wavefunction as a Feynman integral integrated against an energy-fixing kernel. Using this representation of the wavefunction, we have shown that the wavefunction can be written as a total energy pole, with the amplitude as the residue, plus sub-leading terms which are less divergent as $\omega_T\rightarrow 0$ (and are less UV divergent as well). This allow us to study the analytic structure of the wavefuction. The singularities of the wavefunction can be divided into two sets: amplitude-type singularities which can be mapped to singularities from an amplitude, as well as wavefunction-type singularities which are from the subleading term and are unique to wavefunctions. In particular, we demonstrated that the wavefunction-type singularities are not present for in-in correlators at one loop, and we conjecture that this is true at higher loop order as well. 

There are several future directions that are worth investigating:
\begin{itemize}
    \item It would be useful to have a better way to classify the wavefunction-type singularities. For one loop diagrams, we have found that the wavefunction-type singularities are always total energy branch points. For massive fields this is only present for special kinematics, while for massless fields this singularity is always present. However, for more general diagrams, this is less clear. It would be helpful to have a systematic understanding on how these wavefunction-type singularities emerge, what they look like in general, and whether they only appear in specific kinematics.

    One clue is that the wavefunction-type singularities seem to arise from mixing Feynman propagators with boundary terms. A systematic way of writing down these contributions may lead us to such a classification.
    
    \item The ultimate goal is to understand wavefunction analyticity in cosmological spacetime, not just in flat space. In simple cases, we expect the wavefunction for cosmological spacetime to be related to flat space wavefunction simply by taking derivatives with respect to external energies $\omega$. This changes the energy-fixing kernel, but crucially taking derivatives should not introduce additional singularities. Therefore in such cases, we expect the analytic structure of the wavefunction to remain unchanged: the singularities would be in the same location, however the type of the singularity may be changed. For instance, a logarithmic branch point may become a pole instead. 
    
    It has been proposed that the wavefunction for any FLRW spacetime can be written in terms of an integral of the flat space wavefunction \cite{Arkani-Hamed:2017fdk}. Such a representation is written down for the recursion representation of the wavefunction, where it can be computed by using intersection theory \cite {De:2023xue} or differential equations \cite{BaumannCohomology, Vietnam}. This representation also allow us to study the space of functions which appear in the wavefunction using the symbol technology, and this is studied extensively in de Sitter \cite{Hillman:2019wgh}. It would be interesting to see if we can also write down our integral representation of the wavefunction in a similar way, and study its implications on the analyticity of wavefunctions in general FLRW spacetime.
    
    Since the cosmological KLN theorem is true also for de Sitter \cite{Agui-Salcedo:2023wlq}, it would be helpful to understand whether wavefunction-type singularities are also absent in in-in correlators of de Sitter. If the de Sitter wavefunction can be obtained from taking derivatives of the flat space wavefunction, then this conjecture should continue to hold. However, if the de Sitter wavefunction is written as an integral of the flat space wavefunction, then it may be possible to obtain new total energy branch points just by integrating the $\mathcal{A}_n/\omega_T$ term. It would be interesting to investigate this further. 

    \item It is known that any one loop amplitude can be written in terms of a few master integrals \cite{Chetyrkin:1981qh,Smirnov:2012gma}. It would be useful to understand if this is also true for wavefunction using our representation. Naively such a statement should also be true for wavefunctions as well: we can first decompose the amplitude-like part of the wavefunction into master integrals, and then fix the energies using the energy-fixing kernel. If the amplitude-like part has no new poles in terms of internal energies $s$, then the wavefunction can easily be written in terms of master integrals as well. 

    \item It would also be useful to understand how to put constraints on the wavefunction based on our understanding of amplitudes. We have already studied the simple example for tree-level exchange, where positivity for amplitude seems to imply positivity in certain kinematic regimes. However, it is not clear how this will work in more complicated cases such as one loop diagrams, and it is most certainly worthwhile to study the implications of amplitude positivity on the wavefunction.
\end{itemize}

\section*{Acknowledgement}
I would like to thank Enrico Pajer, Scott Melville and Santiago Agüí Salcedo for detailed comments for the draft, as well as Guilherme Pimentel, Arthur Lipstein and Miguel R. Correia for valuable discussions. The author is supported by the Croucher Cambridge International Scholarship.

\appendix

\section{Discontinuity for loop diagrams}\label{app:a}
Let us consider a simple example of a one loop diagram: the two site loop (see figure \ref{fig:2siteclean}). The amplitude representation is given by \eqref{ARpsi2}.

Now we make use of the following fact:
\begin{equation}
    \text{Disc}(f(\omega)g(\omega)h(
    \omega)
    )=[\text{Disc}f(\omega)]g(\omega)h(
    \omega)+f^\ast(-\omega)[\text{Disc}g(\omega)]h(
    \omega)+f^\ast(-\omega)g^\ast(-\omega)[\text{Disc}h(\omega)].
\end{equation}
In addition, we have:
\begin{equation}
    \text{Disc}f(\omega)=2f(\omega)\rightarrow -f^\ast(-\omega)=f(\omega).
\end{equation}
Then we get:
\begin{align*}
    \text{Disc}\psi_2&=\int \frac{ds_1}{2\pi i}\frac{ds_2}{2\pi i}\left[ (\text{Disc}\tilde{D}_1)\tilde{D}_2+\tilde{D}^\ast_1(-\omega_1)(\text{Disc}\tilde{D}_2)\right]\int_{\bfp}\frac{1}{s_1^2-\Omega_{p1}^2}\frac{1}{s_2^2-\Omega_{p2}^2}\\&+\int \frac{ds_1}{2\pi i}\frac{ds_2}{2\pi i}\tilde{D}^\ast_1(-\omega_1)\tilde{D}^\ast_2(-\omega_2)\text{Im}\int_{\bfp}\frac{1}{s_1^2-\Omega_{p1}^2}\frac{1}{s_2^2-\Omega_{p2}^2}
    \\&=\int \frac{ds_1}{2\pi i}\frac{ds_2}{2\pi i}\left[ (2\tilde{D}_1)\tilde{D}_2-\tilde{D}_1(2\tilde{D}_2)\right]\int_{\bfp}\frac{1}{s_1^2-\Omega_{p1}^2}\frac{1}{s_2^2-\Omega_{p2}^2}\\&+\int \frac{ds_1}{2\pi i}\frac{ds_2}{2\pi i}(-\tilde{D}_1)(-\tilde{D}_2)\text{Im}\int_{\bfp}\frac{1}{s_1^2-\Omega_{p1}^2}\frac{1}{s_2^2-\Omega_{p2}^2}\\
    &=\int \frac{ds_1}{2\pi i}\frac{ds_2}{2\pi i}\tilde{D}_1\tilde{D}_2\text{Im}\int_{\bfp}\frac{1}{s_1^2-\Omega_{p1}^2}\frac{1}{s_2^2-\Omega_{p2}^2}.
\end{align*}
Once again we only need to take the imaginary part for the amplitude part of the wavefunction. However in general the disc of a loop diagram is not so simple: for example, if we consider an $n$ site loop (where $n$ is an odd number) we will also need the disc of the energy-fixing kernel. As an example consider the one site loop (figure \ref{fig:1site}). The amplitude representation of this diagram is given by:
\begin{equation}
    \psi_1=\int\frac{ds}{2\pi i}\tilde{D}\int_{\bfp}\frac{1}{s^2-\Omega_p^2},
\end{equation}
where the energy-fixing kernel is:
\begin{equation}
    \tilde{D}=\frac{1}{\omega+2s-i\epsilon}-\frac{2}{\omega-i\epsilon}+\frac{1}{\omega-2s-i\epsilon}.
\end{equation}
The disc of $\psi_1$ is:
\begin{align}
    \text{Disc}\psi_1=&\int\frac{ds}{2\pi i}\text{Disc}\tilde{D}\int_{\bfp}\frac{1}{s^2-\Omega_p^2}+\int\frac{ds}{2\pi i}\tilde{D}\text{Im}\int_{\bfp}\frac{1}{s^2-\Omega_p^2}\\
    &=2\int\frac{ds}{2\pi i}\tilde{D}\int_{\bfp}\frac{1}{s^2-\Omega_p^2}-\int\frac{ds}{2\pi i}\tilde{D}\int_{\bfp}(2\pi i\delta(s^2-\Omega_p^2)).
\end{align}
Doing the $s$ integral gives:
\begin{equation}
    \text{Disc}\psi_1=\int_{\bfp}\frac{1}{2\Omega_p}\left[\frac{1}{\omega+2\Omega_p}-\frac{1}{\omega-2\Omega_p}\right]=\int_{\bfp}\frac{1}{2\Omega_p}\Disc{\frac{1}{\omega+2\Omega_p}}{s}.
\end{equation}
This is the usual result from the cutting rules. 
\pagebreak
\section{Amplitude representation for the wavefunction at two loop}\label{app:b}
Here we show that the wavefunction can be written in the form of \eqref{amprepn} for a simple two loop example. Consider the two site two loop diagram (see figure \ref{fig:2site2loop}):
\begin{figure}[h!]
    \centering
    \includegraphics[scale=1.3]{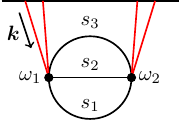}
    \caption{Two site two loop diagram}
    \label{fig:2site2loop}
\end{figure}

The wavefunction is:
\begin{equation}
    \psi_2=\frac{1}{8}\int\frac{ds_1}{2\pi i}\frac{ds_2}{2\pi i}\frac{ds_3}{2\pi i} \tilde{D}_1\tilde{D}_2\int_{\bfp_1,\bfp_2}\frac{1}{s_1^2-\Omega_{p1}^2}\frac{1}{s_2^2-\Omega_{p2}^2}\frac{1}{s_3^2-\Omega_{p3}^2}.
\end{equation}
Here $\bfp_3=\bfk-\bfp_1-\bfp_2$, and 
\begin{multline}
    \tilde{D}_a=\frac{1}{\omega_a+s_1+s_2+s_3-i\epsilon}-\frac{1}{\omega_a-s_1+s_2+s_3-i\epsilon}-\frac{1}{\omega_a+s_1-s_2-s_3-i\epsilon}+\frac{1}{\omega_a-s_1-s_2+s_3-i\epsilon}\\-\frac{1}{\omega_a+s_1+s_2-s_3-i\epsilon}+\frac{1}{\omega_a-s_1+s_2-s_3-i\epsilon}+\frac{1}{\omega_a+s_1-s_2-s_3-i\epsilon}-\frac{1}{\omega_a-s_1-s_2-s_3-i\epsilon}.\label{twoloopD}
\end{multline}
Here $a=1,2$. We can simplify the expression to obtain:
\begin{equation}
    \psi_2=\int\frac{ds_1}{2\pi i}\frac{ds_2}{2\pi i}\frac{ds_3}{2\pi i}\frac{1}{\omega_1+s_1+s_2+s_3-i\epsilon} \tilde{D}_2 \int_{\bfp_1,\bfp_2}\frac{1}{s_1^2-\Omega_{p1}^2}\frac{1}{s_2^2-\Omega_{p2}^2}\frac{1}{s_3^2-\Omega_{p3}^2}.
\end{equation}
Let us do the $s_3$ integral first. The total energy pole comes from the last term in \eqref{twoloopD}, which integrates to:
\begin{equation}
    \psi_2=\int\frac{ds_1}{2\pi i}\frac{ds_2}{2\pi i}\frac{1}{\omega_T}\int_{\bfp_1,\bfp_2}\frac{1}{s_1^2-\Omega_{p1}^2}\frac{1}{s_2^2-\Omega_{p2}^2}\frac{1}{(\omega_2-s_1-s_2)^2-\Omega_{p3}^2}+\dots.
\end{equation}
This is simply the integral for the two loop amplitude, with internal four momentum $p_1=(s_1,\bfp_1)$ and $p_2=(s_2,\bfp_2)$.

It is not hard to see that the remainder terms are less UV divergent. For example, the second last term in \eqref{twoloopD} integrates to:
\begin{equation}
    \int\frac{ds_1}{2\pi i}\frac{ds_2}{2\pi i}\frac{-1}{\omega_T+2s_1}\int_{\bfp_1,\bfp_2}\frac{1}{s_1^2-\Omega_{p1}^2}\frac{1}{s_2^2-\Omega_{p2}^2}\frac{1}{(\omega_2+s_1-s_2)^2-\Omega_{p3}^2}
\end{equation}
It is straightforward to see that after doing another contour integration, this integral is less divergent than the amplitude term.
\bibliographystyle{JHEP}
\bibliography{refs}
\end{document}